\documentclass[prb,aps,amssym,nofootinbib,floatfix,reprint,notitlepage]{revtex4-2} 

\usepackage[dvipsnames]{xcolor}
\usepackage{amsmath,amssymb,bbold,bm}
\usepackage{graphicx}
\usepackage[normalem]{ulem}
\usepackage{esvect}
\usepackage{cancel}
\usepackage{comment}
\usepackage{physics}
\usepackage{kaize_pkg}
\usepackage{bigints}
\usepackage{times}

\usepackage[pdfpagelabels,colorlinks=true,breaklinks=true,linktocpage=true,linkcolor=red,urlcolor=blue,pdfduplex=DuplexFlipLongEdge,citecolor=blue,%
pdfauthor={Kaize Wang, Yang Ge, Yashar Komijani}%
]{hyperref}

\newcommand{\sbraket}[1]{\langle #1 \rangle}
\newcommand{\be}{\begin{equation}}
\newcommand{\ben}{\begin{equation*}}
\newcommand{\ee}{\end{equation}}
\newcommand{\een}{\end{equation*}}
\newcommand{\bs}{\begin{split}}
\newcommand{\es}{\end{split}}
\newcommand{\bmx}{\begin{array}}
\newcommand{\emx}{\end{array}}

\newcommand{\bea}{\begin{eqnarray}}
\newcommand{\bean}{\begin{eqnarray*}}
\newcommand{\eea}{\end{eqnarray}}
\newcommand{\eean}{\end{eqnarray*}}
\newcommand{\dg}{^{\dagger}}
\newcommand{\dn}{^{\vphantom{\dagger}}}

\newcommand{\lr}{\leftrightarrow}

\newcommand{\ua}{\uparrow}
\newcommand{\da}{\downarrow}

\newcommand{\bb}[1]{\mathbb{#1}}

\newcommand{\so}{\qquad\rightarrow\qquad}

\newcommand{\andd}{\qquad\text{and}\qquad}

\newcommand{\eps}{\epsilon}

\newcommand{\veps}{\varepsilon}

\newcommand{\sgn}[1]{{\rm sign}{#1}}
\newcommand{\pref}[1]{(\ref{#1})}

\newcommand{\intinf}[1]{\int_{-\infty}^{+\infty}{#1}}
\newcommand{\intoinf}[1]{\int_{0}^{\infty}{#1}}

\renewcommand{\tr}[1]{{\rm Tr}\Big[ #1 \Big]}

\newcommand{\mat}[1]{\left(\bmx{cc}#1\emx\right)}
\newcommand{\matc}[2]{\left(\bmx{#1}#2\emx\right)}

\newcommand{\bw}[1]{\begin{widetext}}
\newcommand{\ew}[1]{\end{widetext}}

\newcommand{\gray}[1]{}

\newcommand{\nothing}[1]{}

\begin{document}

\title{A Mean-Field Study of Quantum Oscillations in Two-Dimensional Kondo Insulators}
\author{Kaize Wang$^{1,2}$}
\author{Yang Ge$^{2,3}$}
\author{Yashar Komijani\,$^{2*}$}
\affiliation{$^1${Max Planck Institute for the Structure and Dynamics of Matter, Hamburg, 22761, Germany}}
\affiliation{$^2$Department of Physics, University of Cincinnati, Cincinnati, Ohio, 45221, USA}
\affiliation{$^3${Department of Physics and Engineering Physics, Tulane University, New Orleans,  LA 70118, USA}}

\date{\today}
\begin{abstract}
	
Magnetic oscillations in strongly correlated insulating systems have garnered interest due to oscillations seemingly originating from the bulk, despite an anticipated gapped spectrum.	We use the large-$N$ mean-field theory to study the behavior of normal and topological Kondo insulators under a magnetic field. In both cases spinons acquire a charge and hybridize with electrons, producing magnetic oscillations that resemble two-band noninteracting systems. We show that in such band insulators magnetic oscillations are exponentially suppressed at weak magnetic fields.
	{A self-consistent mean-field calculation for the Kondo insulators reveals that the temperature dependence of the oscillations departs from the noninteracting case due to the temperature and magnetic-field dependence of the hybridization, even though mean-field parameters remain homogeneous at low fields.}
	Larger magnetic field results in the Kondo breakdown, where the magnetic oscillation is solely due to the decoupled conduction electrons. These findings offer new insights into the magnetic properties of Kondo insulators, with implications for interpreting experimental results in heavy fermion materials like SmB$_6$.
\end{abstract}
\maketitle
\section{Introduction}

Strongly correlated systems exhibit a fascinating interplay between the quantum behavior of electrons and their mutual Coulomb interaction, giving rise to novel patterns of quantum entanglement and a myriad of largely unexplored properties.

A valuable tool for unveiling the intricacies of the Fermi surface (FS) in materials is the de Haas–van Alphen (dHvA) oscillation. Marked by $1/B$-periodic magnetization oscillations, these phenomena offer profound insights into the FS's shape and volume. Originating from electrons traversing the FS perpendicular to a magnetic field \cite{Shoenberg1984}, dHvA oscillations provide crucial information about the electron density in the system \cite{Luttinger1960} and the effective mass of charge carriers \cite{lifshitz1956}.

While traditionally applied to Fermi liquid metals, dHvA oscillations can be employed to explore the FS in systems beyond Fermi liquid theory. This is particularly relevant in the study of heavy fermion materials, a class of intermetallic compounds containing rare earth or actinide elements. They are theoretically modeled by the Kondo lattice model, in which a lattice of local moments created by correlated electrons is coupled to a lattice of weakly correlated conduction electrons \cite{Coleman2015}. 
As a result of Kondo screening, the FS is expanded to accommodate the density of local moments \cite{Oshikawa00}, and even drives a metal to an insulator in the case of Kondo insulators. Magnetic oscillations have been used to probe such FS expansion across a Kondo breakdown (KBD) transition in heavy fermions \cite{Shishido05} as well as the large effective mass of the carriers. 

Topological Kondo insulators (TKIs) represent a unique class of materials where strong correlation coexists with band topology \cite{dzero2010topological,dzero2012,dzero2012spntopo,Coleman2015,dzero2016}. The heavy-fermion material SmB$_6$ serves as a prime example, exhibiting surface transport responsible for the saturation of resistivity at low temperature  \cite{Wolgast2013,Kim2013}. 

Magnetic oscillations in topological Kondo insulators and in particular SmB$_6$ have attracted immense interest due to a series of experiments in bulk insulators that nevertheless indicate oscillations originating from an apparently bulk Fermi surface \cite{Tan2015}. While oscillations could be expected from the metallic surface of a topological insulator \cite{Wang2010,Alexandrov2015}, the experimental data suggest that these oscillations primarily originate from the bulk. Moreover, optical conductivity measurements of SmB$_6$ hint at its dual nature, acting as a dc insulator but potentially exhibiting ac conductivity \cite{Laurita2016}. Despite challenges related to sample quality and growth conditions \cite{Rosa2020, Li2020}, there has been consistent indications of bulk Fermi surface oscillations \cite{Hartstein2020}, extending to other interacting systems such as the Kondo insulator YbB$_{12}$ \cite{Liu2022} and kagome Mott insulators \cite{Zheng2023}. 

Theoretical efforts to explain this enigma has invoked gap modulations by the magnetic field \cite{Lee2021}, in-gap states due to disorder from impurities \cite{Shen2018} or donors \cite{Skinner2019}, as well as collective modes like excitons \cite{Knolle2017,Spurrier2019,Allocca2022}, with partial experimental success \cite{Fuhrman2015}. In addition, more exotic states have been postulated, including a Majorana Fermi surface \cite{Baskaran2015,Varma2020}, failed superconductors \cite{Erten2017} and neutral fermionic composite excitons \cite{Chowdhury2018,Sodemann2018}. The theoretical controversy stems from the fact that, while magnetic oscillations in the cases of one or two Kondo impurities can be computed exactly \cite{Chou2022}, no such solutions exist in the case of a lattice.

In parallel to entertaining exotic scenarios from a theoretical point of view, it is imperative to carefully investigate to what extent the conventional wisdom about heavy-fermions admits magnetic oscillations. Such wisdom mostly comes from the large-$N$ static mean-field (MF) theory, according to which the spins are fractionalized into fermionic spinons, which hybridize with conduction electrons giving rise to the insulating gap \cite{Coleman2015}. 
Instead of the electromagnetic charge, the spinons carry a charge with respect to an internal gauge field. In presence of a nonzero hybridization, the difference between the internal and external gauge fields is Higgsed \cite{Senthil2003,Coleman05m,Wugalter20}, so that spinons respond coherently to the external electromagnetic field and consequently contribute to the FS. 

In this paper, we use the static large-$N$ theory to study both normal Kondo insulators (NKI) and TKIs in two-dimension in presence of a magnetic field by solving the corresponding MF equations self-consistently. The aforementioned Higgs mechanism is also present in a topological Kondo insulator. A self-consistent solution is effectively equivalent to a density-functional calculation for fractionalized particles. For simplicity we neglect Zeeman splitting and only focus on the orbital effect of the magnetic field.

A central issue is whether MF parameters remain homogenous in presence of a magnetic field. Once this homogeneity is established, magnetic oscillations in NKIs and TKIs are not very different from those of normal insulators (NIs) or topological insulators (TIs), respectively. Magnetic oscillations in band insulators has been studied before \cite{zhang2016quantum,Knolle2015,Allocca2022}. We revisit the problem and by extending the Lifshtiz-Kosevitch (LK) theory to insulators, showing that oscillations persist down to zero temperature. Furthermore, in a Kondo insulator MF parameters (in particular the indirect gap) are temperature dependent. This leads to the fact that, in contrast to band insulators or metals described by the Lifshitz-Kosevich (LK) \cite{lifshitz1956} framework, the intensity of magnetic oscillations not only remains undiminished, but also increases as the temperature rises. This trend is eventually interrupted by the KBD which is the fate of all Kondo insulators at very strong magnetic field due to the suppression in the density of states. 

Numerically the problem is challenging, as the systems studied must be large enough to enable threading multiple units of quantum flux throughout the system without approaching a full flux through each unit cell, at which point the Hofstadter physics takes over \cite{Hofstadter1976}. In a TKI this is further complicated due to i) the susceptibility of the system to field-induced Kondo breakdown and ii) the magnetic oscillations are overshadowed by significant Aharonov-Bohm (AB) oscillations coming from the edge states \cite{Peng2009}, if the thermal length is comparable to the size of the system. Therefore, in this paper we limit ourselves to two-dimensional Kondo insulators, leaving the three-dimensional problem for the future. The simplicity in two-dimension is that the two spin-sectors decouple at the MF Hamiltonian level, although they are still coupled in the self-consistency equations.

The outline of the rest of the paper is as follows: In Section II, we provide the microscopic models of NKI and TKI that are studied throughout the paper. Section III contains our analytical results on quantum oscillation including generalization of the LK theory to NI and TI as well as the effective theory of gauge fields for a TKI, both derived here for the first time. In Section IV, we discuss the homogeneity of MF parameters in a magnetic field and present our numerical results on magnetic oscillations and compare it to the modified LK theory. Section V provides a summarizing discussion and open questions. A series of appendices provide further details and proofs for the statements made in the paper. 

\begin{figure}[ht]
	\includegraphics[width=\linewidth]{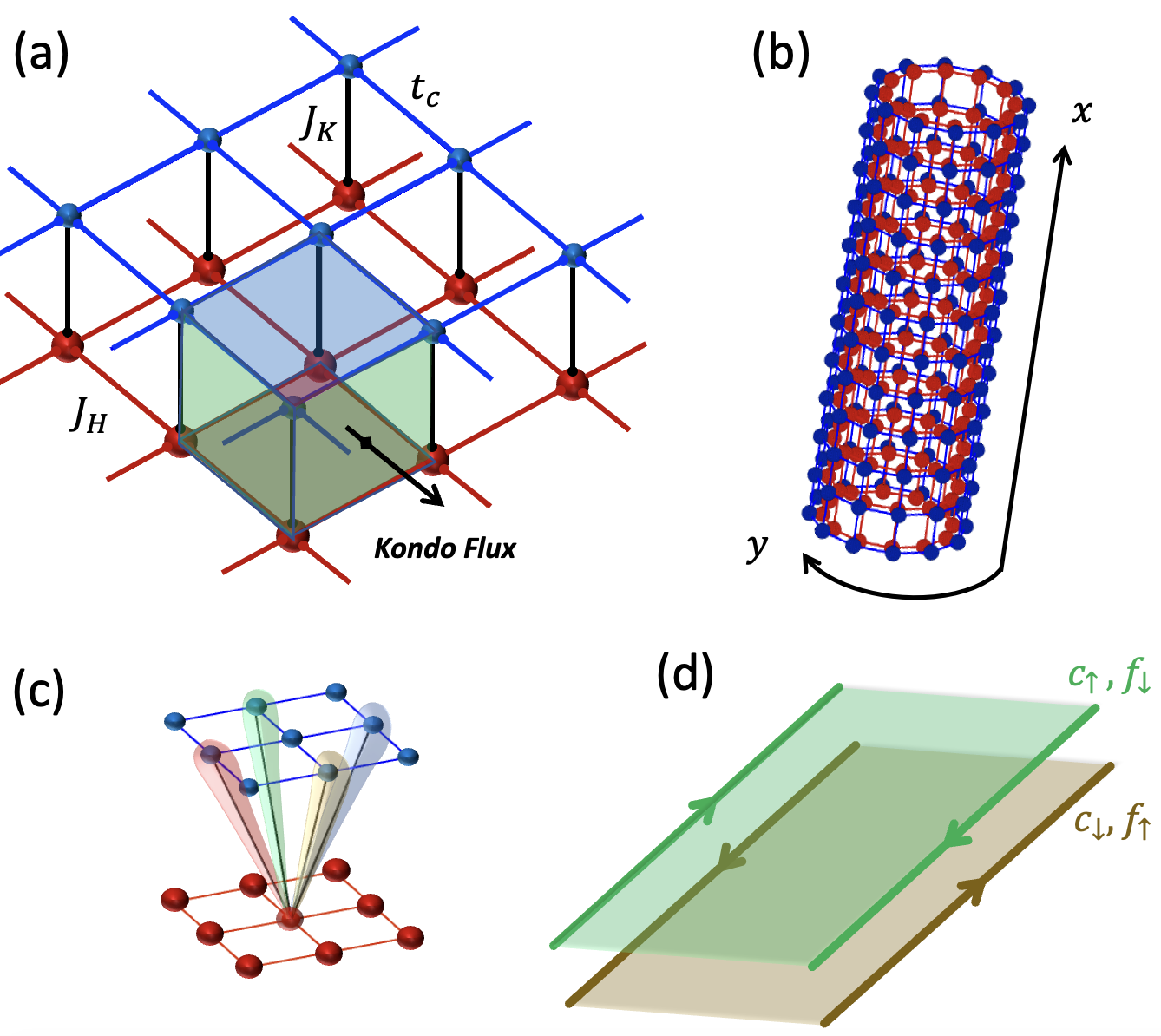}
	\caption{\small {(a) A schematic of the 2D normal Kondo lattice. The local moments made of $f$ electrons (red) interact with each other with Heisenberg coupling $J_H$ and with conduction $c$ electrons (blue) with Kondo coupling $J_K$. The Kondo flux is a mean-field notion corresponding to the product of all phases around a vertical plaquette. Strong Kondo coupling is associated with Kondo flux repulsion in all vertical plaquettes, leading to a Meisner effect that forces the flux piercing the $c$-electron layer to pass through the $f$-electron layer as well. (b) The cylindrical geometry used for the numerical analysis of magnetic oscillations. (c) In a 2D topological Kondo lattice the local moments are coupled to a linear combination of the neighboring conduction electrons which has a definite phase vorticity. (d) The 2D TKI decouples into two spin sectors. Each sector is insulating in the bulk and has edge states with admixtures of $f$ and $c$ electrons with different spins, running in opposite directions.}}\label{fig1}
\end{figure}

\section{Model for Kondo insulators}

The Kondo lattice (KL) model describes conduction electrons coupled to a lattice of localized SU(2) spins (Fig.\,\ref{fig1}a,b), and is mathematically described by the Hamiltonian
\begin{equation}
H = H_c +  \sum_{{\bf r}\boldsymbol{\delta}} J_H^{\boldsymbol{\delta}}\vec{S}_\mathbf{r} \cdot \vec{S}_{{\bf r}+\boldsymbol{\delta}} +  \sum_{\mathbf{r} } J_K^{\boldsymbol{\delta}\boldsymbol{\delta}'}\vec{S}_\mathbf{r} \cdot \vec{s}_{\mathbf{r}+{\boldsymbol{\delta}},\mathbf{r}+{\boldsymbol{\delta}}'}. \label{eq:H}
\end{equation}
We take the (bold-faced) symbol ${\bf r}$ to cover a two-dimensional square lattice and $\boldsymbol{\delta}$ the displacement to all the nearest-neighbors (NN). The conduction electrons hop around according to $H_c=-\sum_{{\bf r}\boldsymbol{\delta}\alpha}(t_c c\dg_{\mathbf{r},\alpha} c\dn_{\bf{r}+\boldsymbol{\delta},\alpha} + \mathrm{h.c.}) -\mu \sum_{\mathbf{r}\alpha} c\dg_{\mathbf{r},\alpha} c\dn_{\mathbf{r},\alpha}$  where $\alpha$ is the spin index. We have considered a generalized version of the KL model with non-local Kondo interaction that can incorporate both normal and topological Kondo insulators (see below). The second term in Eq.\,\eqref{eq:H} represents the Heisenberg coupling between localized spins, and the last term with $\vec{s}_{\bf r,r'}\equiv \frac{1}{2}\sum_{\alpha\beta} c\dg_{\mathbf{r},\alpha} \vec{\sigma}\dn_{\alpha\beta} c\dn_{\mathbf{r'},\beta}$ governs the Kondo screening of spins by the conduction electrons, with $\vec{\sigma}$ denoting the Pauli matrices. Hermicity requires that the Heisenberg couplings $J_H^{\boldsymbol{\delta}}$ are real, and $J_K^{\boldsymbol{\delta}\boldsymbol{\delta}'}$ constitute a hermitian matrix $\bb J_K$ of the Kondo couplings.

We represent localized spins by Abrikosov fermions $\vec{S}_\mathbf{r}= \frac{1}{2} f\dg_{\mathbf{r}\alpha}\vec{\sigma}\dn_{\alpha\beta} f\dn_{\mathbf{r}\beta}$ with a population fixed by the size of the spin \cite{abrikosov1965,Coleman2015}. The problem can be casted in the large-$N$ limit by generalizing SU(2) symmetry group to SU($N$) which favors decoupling the interaction terms in certain channels and reduces the problem to a quadratic Hamiltonian \cite{Coleman2015}. The Heisenberg term becomes
\begin{eqnarray}
	H_f &=& \sum_{{\bf r}\boldsymbol{\delta} }\frac{|t_f^{\mathbf{r} \boldsymbol{\delta}}|^2}{J_H} + \sum_{\mathbf{r} \boldsymbol{\delta},\alpha}\left(t_f^{\mathbf{r}\boldsymbol{\delta}} f\dg_{\mathbf{r}\alpha}f\dn_{\mathbf{r}+\boldsymbol{\delta},\alpha}+\mathrm{h.c.}\right) \nonumber\\
	& & + \sum_{\mathbf{r}} \lambda_\mathbf{r} \Big(\sum_\alpha f\dg_{\mathbf{r}\alpha} f\dn_{\mathbf{r}\alpha} - Q\Big),  \label{eq:HS}
\end{eqnarray}
where $\lambda_\mathbf{r}$ enforces the spin size constraint $n_f = Q$ at each site. The Kondo term becomes the hybridization $V$ between an $f$-electron at the site $\mathbf{r}$ and its neighboring $c$-electrons at the site $\mathbf{r}+\boldsymbol{\delta}$:
\bea
	H_V &=& \sum_{\mathbf{r}\boldsymbol{\delta}\boldsymbol{\delta}',\alpha\beta}V^{\mathbf{r}\boldsymbol{\delta}*}_{\alpha\beta}(\bb J_K^{-1})_{\boldsymbol{\delta}\boldsymbol{\delta}'}V^{\mathbf{r}\boldsymbol{\delta}'}_{\alpha\beta}\nonumber\\
	&&\hspace{2cm}+ \sum_{\mathbf{r}\boldsymbol{\delta},\alpha\beta}\left(V^{\mathbf{r}\boldsymbol{\delta}}_{\alpha\beta}c\dg_{\mathbf{r}+\boldsymbol{\delta},\alpha}f\dn_{\mathbf{r},\beta}+\mathrm{h.c.}\right)\!. \label{eq:HK}
\eea
We introduce magnetic field by the Peierls substitution on $t_c$ hopping. Despite similar forms of $H_c$ and $H_f$, an external magnetic field only adds a hopping phase to $t_c$, while the $f$-electrons have a separate internal $\mathrm{U}(1)$ gauge freedom.
\bea
f_{{\bf r}\alpha}\to e^{i\varphi^{\rm in}_{\bf r}},\quad t_f^{\bf r\alpha}\to t_f^{\bf r\alpha}e^{i(\varphi^{\rm in}_{\bf r}-\varphi^{\rm in}_{\bf r+\delta})}, \quad V_{\alpha\beta}^{\bf r\boldsymbol\delta}\to V_{\alpha\beta}^{\bf r\boldsymbol\delta}e^{i\varphi_{\bf r+\boldsymbol\delta}^{\rm in}}\nonumber.
\eea
Note that $V^{\mathbf{r}\boldsymbol{\delta}}_{\alpha\beta}$ also carries charge w.r.t.\ external electromagnetism and which leads to the Higgs mechanism when $V^{\mathbf{r}\boldsymbol{\delta}}_{\alpha\beta}$ condenses. As we will see shortly, the form of $V^{\mathbf{r}\boldsymbol{\delta}}_{\alpha\beta}$ affects the topology of a Kondo insulator \cite{Coleman2015}. A large-N treatment of the model requires a rescaling $\bb J_K\to J_K/N$ and $J_H\to J_H/N$. Within the large-$N$ MF theory, the free energy of the Kondo system is
\begin{equation}
	F[\lambda,V,t_f] = - T \ln \Tr ( e^{-H[\lambda,V,t_f]/T}),
\end{equation}
where $T$ is temperature. The MF parameters $V$ and $t_f$ are c-numbers that follow mean field equations given by 
$\partial F/\partial V^*=0$ and $\partial F/\partial t_f^* = 0$. The spin size is enforced by $\partial F/\partial \lambda =0$.

\subsection{Normal Kondo Insulators}
For a normal Kondo insulator (NKI), the hybridization is a local spin-singlet $V_{\alpha \beta}^{\mathbf{r}\boldsymbol{\delta}} = V_{\mathbf{r}}{\delta}_{\alpha\beta}{\delta}_{\boldsymbol{\delta}\mathbf{0}}$. The hybridization term takes the form
\begin{equation}
	H_V = \sum_{\mathbf{r}\alpha} \left( V_{\mathbf{r}} c_{\mathbf{r}\alpha}^\dagger f_{\mathbf{r}\alpha} + \mathrm{h.c.} \right)
	+ N\frac{|V_\mathbf{r}|^2}{J_K}.
\end{equation} 
and the MF equations become
\begin{eqnarray}
		V_\mathbf{r} &=& -\frac{J_K}{N} \sum_\alpha \langle f^\dagger_{\mathbf{r}\alpha} c\dn_{\mathbf{r}\alpha} \rangle, \nonumber\\
		t_f^{\mathbf{r}\boldsymbol{\delta}} &=& -\frac{J_H}{N}  \sum_\alpha \langle f\dg_{\mathbf{r}\alpha} 
		f\dn_{{\mathbf{r}+\boldsymbol{\delta}}\alpha} \rangle  ,   \nonumber\\
		Q &=& \sum_\alpha \langle f\dg_{\mathbf{r}\alpha} f\dn_{\mathbf{r}\alpha}\rangle,
		\label{eq:MF}
\end{eqnarray}
where $\boldsymbol{\delta}=a \hat{\mathbf{x}},a \hat{\mathbf{y}}$ are displacements to the NNs of the lattice site at $\mathbf{r}$. From now on we set the lattice constant $a=1$. 

\subsection{Topological Kondo Insulators}

The topological Kondo insulators, first introduced by Ref.\,\cite{dzero2010topological}, is the interacting analog of the time-reversal invariant topological insulators where nontrivial topology arises from band inversion. In the presence of strong spin-orbit coupling, the hybridization that induces band inversion will only preserve the total angular momentum comprised of both spin and orbital angular momenta, while the time-reversal (TR) symmetry relates subspaces with opposite total angular momenta. This is particularly the case if the two orbitals belong to the same atom ($d$ and $f$ orbitals of Sm in SmB$_6$), so that they cannot couple unless the hybridization contains a vorticity to compensate the mismatch in angular momenta [Fig.\,\ref{fig1}(c)]. This can be obtained by decoupling the Kondo interaction in the triplet channel which can also  be cast in the large-$N$ limit \cite{dzero2012spntopo}. We skip these details and rather follow the less rigorous but more insightful route \cite{dzero2010topological} of starting from the MF Hamiltonian and figuring out the Kondo coupling matrix afterwards.

The simple model we consider here mimics a 2D reduction of the Bernevig-Hugh-Zhang model, whose full Hamiltonian includes different spin sectors of both $c$ and $f$ electrons \cite{bernevig2006,dzero2010topological,dzero2012,dzero2012spntopo,Coleman2015,dzero2016}.  In momentum space, it is
\begin{eqnarray}
	\label{eq:HTKI}
	H = \sum_\mathbf{k} \begin{pmatrix}
		c_{\mathbf{k}}\\
		f_{\mathbf{k}}
	\end{pmatrix}^\dagger 
	\begin{pmatrix}
		\varepsilon_c \mathbb{1} & V \vec{d}_\mathbf{k}\cdot \vec{\sigma}  \\
		V^*  \vec{d}_\mathbf{k}\cdot \vec{\sigma}   & \varepsilon_f \mathbb{1} 
	\end{pmatrix}
	\begin{pmatrix}
		c_{\mathbf{k}}\\
		f_{\mathbf{k}}
	\end{pmatrix},
\end{eqnarray}
where
\begin{eqnarray}
	\varepsilon_c&=&-2t_c (\cos k_x + \cos k_y) - \mu, \nonumber\\
	\varepsilon_f&=&-2t_f (\cos k_x  + \cos k_y) + \lambda, \nonumber\\
	\vec{d}_\mathbf{k} & = &(\sin k_x ,\sin k_y,0), \quad \vec{\sigma} = (\sigma^x,\sigma^y,\sigma^z).
\end{eqnarray}
The Pauli matrices $\vec{\sigma}$ acts on the spin space. We also define the spinors $c_\mathbf{k} = (c_{\mathbf{k \uparrow}},c_{\mathbf{k}\downarrow})^T$, $f_{\mathbf{k}} = (f_{\mathbf{k}\uparrow},f_{\mathbf{k}\downarrow})^T$.


In this 2D version, the model decouples into two spin sectors made of $(c_{\uparrow},f_{\downarrow})$ and $(c_{\downarrow},f_{\uparrow})$ which are related to each other via time-reversal symmetry. Since each sector is lattice-regularized and the analyses are parallel, we will focus only on one sector in the subsequent discussions, albeit both sectors will be taken into account numerically to satisfy the constraint. Each resulting $2\times2$ MF hamiltonian is equivalent to the Qi-Wu-Zhang (QWZ) model \cite{QWZ2006,bernevig2013,asboth2016} which plays the same role on a square lattice that the Haldane model \cite{Haldane1988} plays on a honeycomb lattice.

Here, the hybridization $V$ takes the form $V_{\alpha \beta}^{\mathbf{r}\boldsymbol{\delta}} = e^{i\tilde\alpha\angle{\boldsymbol{\delta}}} V_\mathbf{r} \delta_{\bar\alpha\beta}$,
where $\tilde{\alpha} = \pm 1$ for $\alpha=\ua,\da$ respectively, $\boldsymbol{\delta} = \pm \hat{\mathbf{x}},\pm \hat{\mathbf{y}}$, and $\angle{\boldsymbol{\delta}}$, increases counterclockwise from 0 for $+\hat{\mathbf{x}}$ to 2$\pi$. 
This vorticity in hybridization phase is shown schematically in Fig.\,\ref{fig1}(c). Alternatively, the hybridization term can be written in terms of
\bea
\breve{c}_{\mathbf{r},+} &\equiv& \frac{1}{2}[c_{\mathbf{r}+\hat{\mathbf{x}},\da} + i c_{\mathbf{r}+\hat{\mathbf{y}},\da} + i^2 c_{\mathbf{r}-\hat{\mathbf{x}},\da}
	+ i^3 c_{\mathbf{r}-\hat{\mathbf{y}},\da}],\\
\breve{c}_{\mathbf{r},-} &\equiv& \frac{1}{2}[c_{\mathbf{r}+\hat{\mathbf{x}},\ua} - i c_{\mathbf{r}+\hat{\mathbf{y}},\ua} + (-i)^2 c_{\mathbf{r}-\hat{\mathbf{x}},\ua}
	+ (-i)^3 c_{\mathbf{r}-\hat{\mathbf{y}},\ua}].\nonumber
\eea
In the continuum limit $\breve c_{+}\sim \partial^+c_\da$ and $\breve c_-\sim\partial^-c_\ua$ where $\partial^\pm=\partial_x\pm i\partial_y$ and therefore $\breve c_{\pm}$ carry orbital angular momentum $\pm1$ and spin $\mp1/2$. Due to spin-orbit coupling, only the total angular momentum is conserved in tunnelling, therefore $\breve c_{+}$ hybridizes with $f_\ua$ and $\breve c_{-}$ hybridizes with $f_\da$:
\begin{equation}
	H_V = \sum_{\mathbf{r},\alpha=\ua\da} \left( V\dn_{\mathbf{r}} \breve{c}_{\mathbf{r},\tilde\alpha}^\dagger 
	f\dn_{\mathbf{r}\alpha} + \mathrm{h.c.} \right) + \frac{4|V_\mathbf{r}|^2}{J_K}.
\end{equation}
{Adding onsite Coulomb repulsion for $f$-electrons (or integrating out the bosonic mode $V$) leads back to the generalized Kondo interaction of Eq.\,\pref{eq:H}.} The MF equations for the TKI then takes a similar form to the NKI case \eqref{eq:MF} with $c_{\mathbf{r},\alpha} \to \breve c_{\mathbf{r},\tilde\alpha}$.

When coupled to $\mathrm{U}(1)$ gauge fields, the hopping term should be modified according to
\begin{equation}
	t_{c,f}^{\mathbf{r}\boldsymbol{\delta}} \rightarrow t_{c,f}^{\mathbf{r}\boldsymbol{\delta}}\exp[i \frac{ e}{\hbar}\int_{\mathbf{r}}^{\mathbf{r}+\boldsymbol{\delta}}
	\mathbf{A}_{c,f}{(\mathbf{s})}\cdot \mathrm{d}\mathbf{s}],
\end{equation}
where $A_c$ is the external electromagnetic field and $A_f$ is the internal gauge field of $f$ electrons.

For the TKI case with non-local, since the hybridization is not onsite, an extra gauge string is needed to maintain gauge invariance. The definition of $\breve c_{\mathbf{r},\tilde\alpha}$ needs to be modified to
\begin{equation}
	\breve{c}_{\mathbf{r},\tilde\alpha} \rightarrow \frac{1}{2}\sum_{\boldsymbol{\delta}} i^{i\tilde\alpha\angle{\boldsymbol{\delta}}} \exp\left[i \frac{ e}{\hbar}\int_{\mathbf{r}}^{\mathbf{r}+\boldsymbol{\delta}}\mathbf{A}_c(\mathbf{s})\cdot \mathrm{d}\mathbf{s}\right]c_{\mathbf{r}+\boldsymbol{\delta},\bar\alpha}.	
\end{equation} 
which transforms covariantly under gauge transformation $c_{\mathbf{r}\alpha}\rightarrow c_{\mathbf{r}\alpha}e^{-i\varphi(\mathbf{r})}$, $\mathbf{A}_c(\mathbf{r})\rightarrow \mathbf{A}_c(\mathbf{r}) + \nabla \varphi(\mathbf{r})$.

\begin{figure}[h!]
\includegraphics[width=0.99\linewidth]{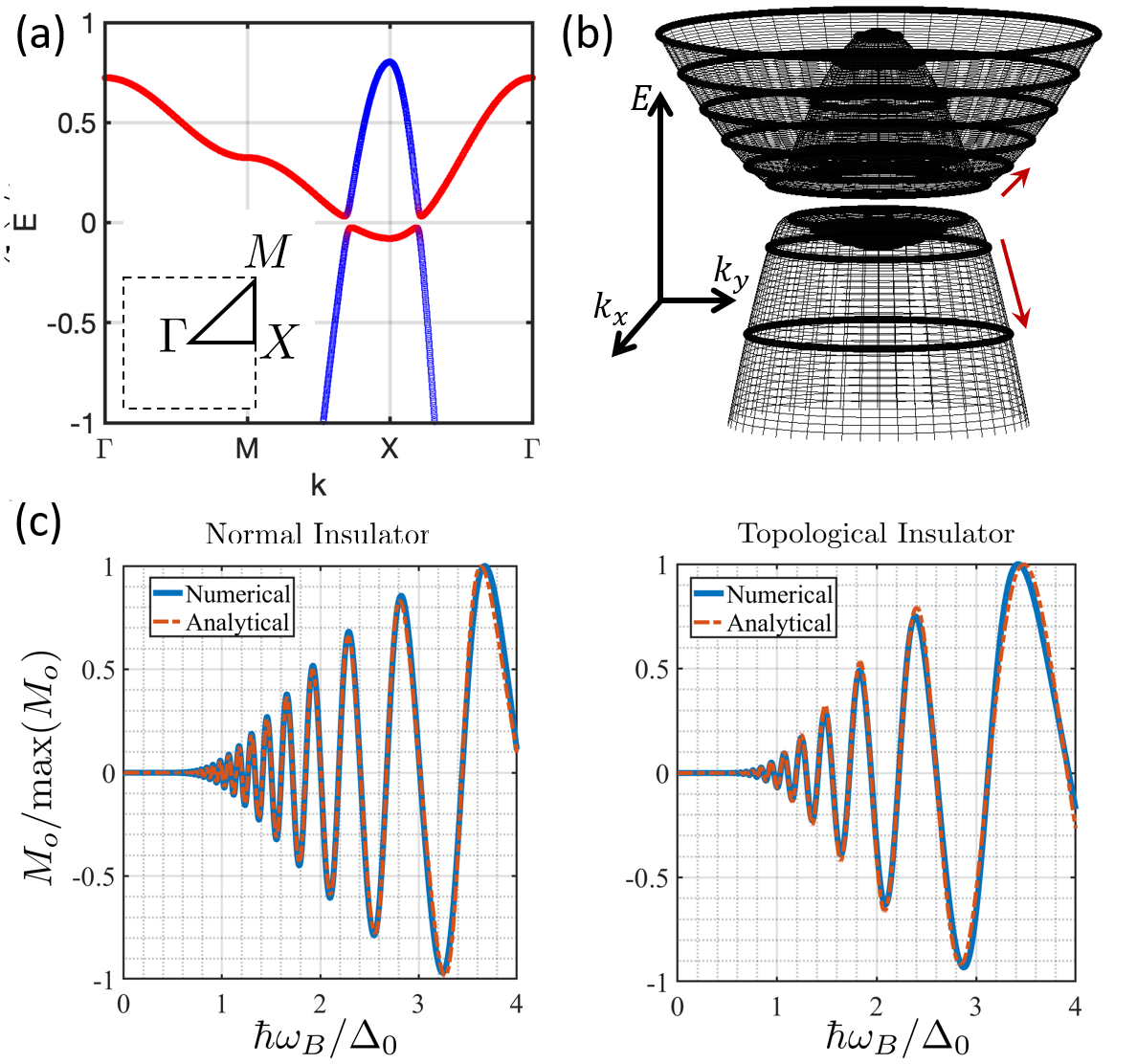}
\caption{\small (a) Dispersion of a TKI. The parameters are $t_f/t_c=-0.1$, $\lambda/t_c=0.3$ $\mu/t_c=3.2$ and $V/t_c=0.05$. Similar parameters have been used for the NKI. (b) Magnetic orbits in a band insulator. The evolution of the Landau levels with increasing magnetic field is marked with red arrows, from the $k=0$  center to larger $k$-s. (c) The oscillating part of magnetization for NI and TI, as a function of normalized magnetic field at $T=0$. Here $\Delta_0 = \Delta(B=0)$, and $\mathrm{max}(M_o)$ is the maximal oscillating amplitude among the plotted field range. Solid blue line: Numerical exact evaluation of Eq.\,\eqref{eq:Fosc}. Dashed red line: Analytic results of Eq.\,\eqref{eq:F_sol}.  For NI, the parameters are $\tilde{\lambda} = 6, \Delta_0 = 0.5$. For TI, the parameters are $\tilde{\lambda} = 6, \Delta_0 = 0.78$.}\label{fig1p5}.
\end{figure}

In the insulating regime, TKI Hamiltonian \eqref{eq:HTKI} is topological when \cite{QWZ2006,asboth2016}
\begin{equation}
	\label{eq:topo}
	0<|\lambda+\mu|<4|t_c+t_f|,
\end{equation}
where the Chern number is given by $\sgn(\lambda+\mu)$. In this phase, the two spin sectors form a time-reversal pair of counter-propagating edge states at the boundary of the 2D system [Fig.\,\ref{fig1}(d)], reflected in a nontrivial $Z_2$ index \cite{dzero2010topological,dzero2012,dzero2012spntopo,Coleman2015,dzero2016,FuKane2007,bernevig2013,Coleman2015,asboth2016} similar to the quantum spin Hall effect. An example of the resulting band structure is shown in Fig.\,\ref{fig1p5}(a).


\section{Analytical Results}
In this section we provide a number of analytical insights into the electromagnetic response of Kondo insulators.
\subsection{Magnetization}
For both types of Kondo insulators, the MF Hamiltonian can be diagonalized with single-particle orbitals. In such an eigenbasis, the Hamiltonian becomes
\begin{equation}
H = \sum_{l} \varepsilon_l \psi_l\dg \psi\dn_l +  N\sum_{\mathbf{r}\Delta}\frac{|t_f^{\mathbf{r} \Delta}|^2}{J_H} + N\sum_\mathbf{r} \frac{|V_\mathbf{r}|^2}{J_K}- \lambda_\mathbf{r} Q ,
\end{equation}
where $l$ is the single-particle orbital index, $\varepsilon\dn_l$ is the energy of the orbital, and $\psi_l$ ($\psi\dg_l$) is the corresponding annihilation (creation) operator. Here we have separated out contributions from $V,t_f$ and $\lambda$, and omitted the spin indices.
The Gibbs free energy is given by
\begin{eqnarray}
	F[B;\lambda,V,t_f] &=& -T\sum_{l} \ln\left[1 + \exp(-\varepsilon_l/T) \right]\\
 & &+ N\sum_{\langle \mathbf{r} \mathbf{r}' \rangle}\frac{|t_f^{\mathbf{r} \mathbf{r}'}|^2}{J_H} + N\sum_\mathbf{r} \frac{|V_\mathbf{r}|^2}{J_K}- \lambda_\mathbf{r} Q.\nonumber
\end{eqnarray}
Magnetization can be extracted from the free energy using
\begin{eqnarray}
	M&=&-\frac{\partial F}{\partial B}\nonumber\\
	&=&-\sum_l  f(\varepsilon_l)\frac{\partial\varepsilon_l}{\partial B}-
	\frac{\partial V}{\partial B}\frac{\partial F}{\partial V}-	\frac{\partial t_f}{\partial B}\frac{\partial F}{\partial t_f}-	\frac{\partial \lambda}{\partial B}\frac{\partial F}{\partial \lambda}\nonumber\\
	&=&-\sum_l  f(\varepsilon_l)\frac{\partial\varepsilon_l}{\partial B}. \label{eq:M}
\end{eqnarray}
All partial derivatives by $V$, $t_f$ and $\lambda$ vanish due to the saddle point conditions. Note that the magnetization still implicitly depends on MF parameters through $\varepsilon_l$'s.

\subsection{Modified Lifshitz-Kosevich formula for normal and topological band insulators  }

The Landau levels (LL) in an insulator are schematically depicted in Fig.\,\ref{fig1p5}(b). As the magnetic field changes, the occupied LLs in the valence band move toward the band edge and then down to negative energies. Therefore, they contribute an oscillatory factor to the free energy which leads to $1/B$ periodic magnetic oscillations at zero temperature. 

The analytical calculation of quantum oscillation, originally derived by LK for a metal, can be extended to the case of an insulator \cite{Grubinskas2018}. We have further extended these calculations to both normal (NI) and topological (TI) band insulators. We assume a continuum model in which in absence of hybridization, the $c$ and $f$ electrons each have quadratic dispersion with effective masses $m_c$ and $m_f$. Here, we only present the results and refer the readers to Appendix \ref{AF} for the detailed derivations. For $T$ much smaller than the hybridization gap $\Delta$, the oscillatory part of the free energy is given by
\be
	\label{eq:Fosc} 
\hspace{-.6em} F = \mathcal{N} \Delta\,\text{Im}  \sum_{\ell=1}^{\infty} \frac{(-1)^{\ell}}{\pi \ell} I_\ell
	\left( {\Delta}/{\hbar \omega_B};x_0 \right) \exp ( 2 \pi i \ell \frac{\tilde{\lambda}}{\hbar \omega_B})\!,
\ee
where $\omega_B= eB/m$, with the reduced mass $m= 2\abs{m_cm_f}/(\abs{m_c}+\abs{m_f})$, is the cyclotron frequency, $\mathcal{N} = BA e/h$ is the Landau level degeneracy, $\tilde{\lambda}$ is related to the chemical potential of the bare $c$ and $f$-electron dispersion. The oscillations' amplitude $I_\ell$ is given by
\bea 
	 I_\ell
	\left( {\Delta}/{\hbar \omega_B};x_0 \right) = \int_{-x_0}^\infty \frac{x}{\sqrt{x^2 + 1}} \exp \left( 2 \pi i \ell \frac{\Delta}{\hbar \omega_B} x\right),\label{eq:LKInt}
\eea 
where $x_0 = (\tilde{\lambda} - \hbar \omega_B / 2)/\Delta$ is roughly the ratio of Fermi energy and the gap size.

The numerical results for both TI and NI are shown in Fig.\,\ref{fig1p5}(c) at $T\ll\Delta$. The oscillation amplitude remains finite as $T \to0$. At low field, there are no oscillations but the amplitude grows when magnetic field is comparable to the gap $\Delta$. This can be expressed compactly under the assumption that the number of occupied Landau levels is large and gap is small, i.e.\ $x_0\to\infty$. In this limit, $F$ can be expressed analytically with modified Bessel function of second kind, $K_1(x)$,
\bea
	\label{eq:F_sol}
	F = 2  \mathcal{N} \Delta \sum_{\ell=1}^\infty \frac{(-1)^\ell}{\pi \ell} 
	K_1\left(2 \pi \ell \frac{\Delta}{\hbar \omega_B} \right) \cos\left( 2 \pi \ell \frac{\tilde{\lambda}}{\hbar \omega_B}\right)\!.\; 
\eea 	
The asymptotic behavior of $K_1(x)\sim x^{-1/2} e^{-x}$ as $x \to \infty $ results in an exponential suppression of oscillations at small field. {On the other hand, as $x \to 0$, $K_1(x) \sim 1/x$, generating a finite amplitude of oscillation when magnetic field increases. In this limit our result approaches the LK theory of un-hybridized metallic bands.}

Finally, we relate the formal parameters in Eq.\,\eqref{eq:Fosc} to the those in the NI and TI Hamiltonians. For NI,  $\Delta=V$ is the hybridization and $\tilde{\lambda}=\lambda$ is the chemical potential difference of bare $c$ and $f$ electrons. The oscillatory argument can be written as $2\pi\ell {{\cal A}_F}/{heB}$ where ${\cal A}_F$ is the FS area enclosed by the contour where the dispersion of $c$ and $f$ will intersect if the hybridization is turned off. For TI, the parameters are $\tilde{\lambda}= \lambda - m V^2$ and $\Delta = \sqrt{m V^2(m V^2 - \hbar \omega_B + 2 \tilde{\lambda})}$.

\subsection{Effective theory of gauge fields}

The electromagnetic response in Kondo lattice systems at low energy and long wavelength is marked by the locking of the internal gauge $A_f$ to the external applied field $A_c$ \cite{Senthil2003,Coleman05m,Coleman2015,Wugalter20}. The locking manifests itself in the effective gauge action after integrating out the fermions $c$ and $f$. Below we review the derivation of this action for NKI and re-derive it for a TKI.

\subsubsection{Low energy action and the Chern-Simons term}
A continuum limit description of the problem can be obtained by focusing on the modes near the chemical potential. While the low energy modes in an NKI are concentrated at the band crossing, they are more complicated in a TKI. In this case, Equation \eqref{eq:HTKI} implies that the conduction band bottom and valence band top can appear at generic momenta. However, the $k$-points where the hybridization vanish, namely at  $\Gamma$, $M$, $X$ and $Y$, are more significant as they are the sources of the Berry's phaser and the Chern-Simons (CS) term in the effective action.
These are also the low energy momenta close to topological phase transitions in a TKI or the QWZ model. Near topological phase boundaries, we may carry out low energy expansions at these critical momenta, so that $V\sin k \sim V k$ and $t \cos k \sim t - k^2/2m $, where $m=1/t a^2$. This leads to Dirac cones at $X$ and $Y$ when $|\lambda+\mu|\sim0$, and at $\Gamma$ or $M$ when $\lambda+\mu\sim \pm 4|t_c+t_f|$.


In each spin sector, the low energy modes at $\Gamma$, $M$, $X$ and $Y$ contribute to half of an integral CS term \cite{Ma2018,bernevig2013}. In the topological regime, the terms at $X$ and $Y$ adds up, while those from $\Gamma$ and $M$ cancels, dictated by the sign of the Dirac mass at each momentum. The resulting (integral) CS terms in the two spin sectors carry opposite signs and cancel each other due to the time-reversal symmetry. Thus, there is no net CS term in the effective gauge theory.

Our discussion on the CS term so far has assumed a single $\mathrm{U}(1)$ gauge field coupled to all the fermions, as in a noninteracting TI. Next, we will show that the gauge locking in Kondo insulators unifies $A_c$ and $A_f$. Therefore, this discussion also applies to topological Kondo insulators.

\subsubsection{Effective gauge theory and the Higgs term}

Now we turn to the effective gauge theory that locks $A_f$ to $A_c$. As discussed in the preceding subsection, only $k$-points where the hybridization vanish are relevant.
For NKI, since the hybridization is constant in $k$-space, it suffices to the low energy mode near $\Gamma$ point, while for TKI one needs to consider all of $\Gamma$, $M$, $X$ and $Y$.
Since calculations at these $k$-points are similar, we only show the derivation at $\Gamma$.

The Hamiltonian density obtained from $\Gamma$ point expansion is $\mathcal{H}(x) = \mathcal{H}_0(x) + 
\mathcal{H}_{V}(x)$ where $\mathcal{H}_V(x)$ is the continuum version ofEq.\,\eqref{eq:HK} and $\mathcal{H}_0(x)$ is given by
\begin{eqnarray}
	\mathcal{H}_0(x)  & = &
	c^\dagger(x) \left[ \frac{(\mathbf{p} - e \mathbf{A}_c)^2}{2 m_c} + ie A_c^\tau- \mu \right] c(x) \nonumber \\
	&&+ f^\dagger(x) \left[ \frac{(\mathbf{p} -  \mathbf{A}_f)^2}{2 m_f} + i A_f^\tau + \lambda \right] f(x) ,
\end{eqnarray}
where $A^\tau$ is the temporal component of the gauge potential.
Setting $e=1$, the Lagrangian density is given by
\begin{eqnarray}
	\mathcal{L} & = &\bar{c} \left[(\partial_\tau - \mu) + i A_c^{\tau} - \frac{1}{2 m_c} 
	(\nabla + i \mathbf{A}_c)^2\right] c \nonumber \\
	&&+\bar{f} \left[(\partial_\tau + \lambda) + i A_f^{\tau} - \frac{1}{2 m_f} 
	(\nabla + i \mathbf{A}_f)^2\right] f \nonumber \\
	&&+ c^\dagger \hat{V} f + f^\dagger \hat{V}^\dagger c + i A_c^\tau n_c + i A_f^\tau Q. 
	\label{eq:L}
\end{eqnarray}
Here $\hat V=V$ for an NKI and $\hat V=iV\partial^\pm$ for the two sectors of a TKI. Apart from hybridization, two additional terms are included above. The term $i A_\tau^c n_c$ is the coupling of the fluctuations in the electromagnetic potential to the positive charge density of the ionic background $n_c$. The other term $i A_f^\tau Q$ comes from the spin size constraint when introducing the Abrikosov fermions \eqref{eq:HS}. 

The effective action for the gauge fields $S_{\mathrm{eff}}[A_c,A_f]$
is obtained by integrating out the fermions, as detailed in Appendix \ref{SE}.
To one-loop, the effective actions of both NKI and TKI take the form of mass terms for the gauge fields,
\begin{equation}
	\label{eq:Seff}
 \frac{S_{\mathrm{eff}}}{N}\!=\!\!\int \! \mathrm{d}^3 x \Gamma \big[(A^{\tau}_c - A^\tau_f + \partial_\tau \varphi)^2 + v_{\Gamma}^2 (\mathbf{A}_c - \mathbf{A}_f+ \nabla \varphi)^2\big],
\end{equation}
where $\varphi \equiv \arg V$ is the hybridization phase, and $\Gamma$ and $v_\Gamma$ are the stiffness and velocity coefficients, respectively.

In an NKI, this result has been derived in the continuum limit and in presence of a UV cut off for the conduction electrons in Refs.\,\onlinecite{Senthil2003,Coleman05m,Wugalter20}. Generalization to the TKI is more subtle and requires the conduction electron bandwidth to be incorporated in a gauge-invariant fashion. This is accomplished by a Pauli-Villars regularization of the conduction electrons  (see Appendix \ref{SE}). The stiffness and velocity thus obtained are
\begin{eqnarray}
	\Gamma &=& \frac{1}{ \pi} \left(\frac{1}{m_c} + \frac{1}{|m_f|}\right)^{-1},\nonumber \\
	v_\Gamma &\sim&\frac{D}{m_c+|m_f|},
\end{eqnarray}
where $D$ is the bandwidth of the conduction electron, identified with the energy cutoff in the Pauli-Villars regulator.

Such an effective action locks the internal $\mathrm{U}(1)$ gauge field $A_f$ to the external gauge field $A_c$. Consequently, the Meissner effect develops and expels the difference field $A^\mu_{c}-A^\mu_f$ from the system (see below). Numerical results in Sec.\ \ref{Sec2} and Appendix \ref{Solver} are consistent with the Kondo flux repulsion and the Anderson-Higgs mechanism both in NKI and TKI. 

Finally, under strong magnetic field $\hbar \omega_B \gg J_K$, Kondo breakdown $(V\to 0)$ occurs and the gauges are no longer locked together.  With gauge-locking established, we also conclude that the CS term in each sector is exactly that of a normal or topological insulator, which leads to no net CS term.

\subsubsection{Kondo flux repulsion}
Within the MF theory, the Higgs term appears as a Meissner effect. The ensuing Kondo flux repulsion is reflected in the phase on three out of four edges of any \emph{Kondo plaquette}, such as the NN pairs of $c$ and $f$ electrons depicted in Fig.\,\ref{fig1}(a), which are determined self-consistently. The Kondo plaquette is diagonalized in Appendix \ref{sec:paquette}. It is shown that the energy is minimized if the product of all hopping amplitudes around the plaquette is real and negative, meaning that the Kondo flux is $\pi$. Consequently, when hybridizations are taken to be real and positive, spinon hopping will acquire the same phase as the electron hopping, up to a sign. Therefore, an external field incorporated via Peierls substitution for the conduction electrons will be imprinted on the internal gauge field of spinons. This agrees with our numerical results in the next section.

\section{Numerical Results}\label{Sec2}

Here, we present our numerical solution to the self-consistent MF equations, in both open boundary condition and cylindrical geometries [Fig.\,\ref{fig1}(a,b)] similar to the calculation done in \cite{Rebecca21}. First, we demonstrate the gauge locking effect in Kondo insulators and the homogeneity of the MF parameters in the open geometry, before the KBD is discussed. Then, the quantum oscillation at low $T$ and KBD at high $T$ are studied in the cylindrical geometry.

\subsection{Homogeneity of MF parameters and gauge locking \label{sec:homo}}

Mean-field treatments of Kondo insulators usually assume uniform MF parameters. This may not be true in interacting systems under a magnetic field, even if the field is uniform. A well-known example is the vortex formation in type-II superconductors \cite{abrikosov1957}. There has been recent proposals for such nonuniform states in heavy-fermion systems \cite{Koenig2023}. Therefore, we first examine the spatial homogeneity of Kondo insulators under a uniform magnetic field. 

Bulk-homogeneity in gapped systems is closely linked to the gauge invariance. Let $\ket{\psi}$ be a groundstate of the MF Hamiltonian $\hat{H}\left[\mathbf{A}\right]=\frac{1}{2m}\left[\hat{\mathbf{p}}-e\mathbf{A}(\hat{\mathbf{r}})\right]^2$ where $\mathbf{A}$ gives a uniform magnetic field. In the bulk, a spatially shifted state $e^{-i\hat{\mathbf{p}}\cdot\mathbf{r_0}}\ket{\psi}$ is the eigenstate of a Hamiltonian with a shifted vector potential, i.e., $\mathbf{A}'(\mathbf{r})=\mathbf{A}(\mathbf{r}+\mathbf{r_0})$. Since $\nabla \times \mathbf{A}'=\nabla \times \mathbf{A}$, this amounts to a gauge transformation. We assume the spectrum to be gapped and the ground state to be non-degenerate. In this case, the spatial shift does not affect charge density or any other observables, nor does it change the ground state.
However, the independent gauge invariance of $\mathbf{A}_c$ and $\mathbf{A}_f$ is broken in a Kondo insulator, and the Kondo hybridization, carrying charge with respect to the two, can in principle vary in the bulk.

\begin{figure}
	\includegraphics[width=\linewidth]{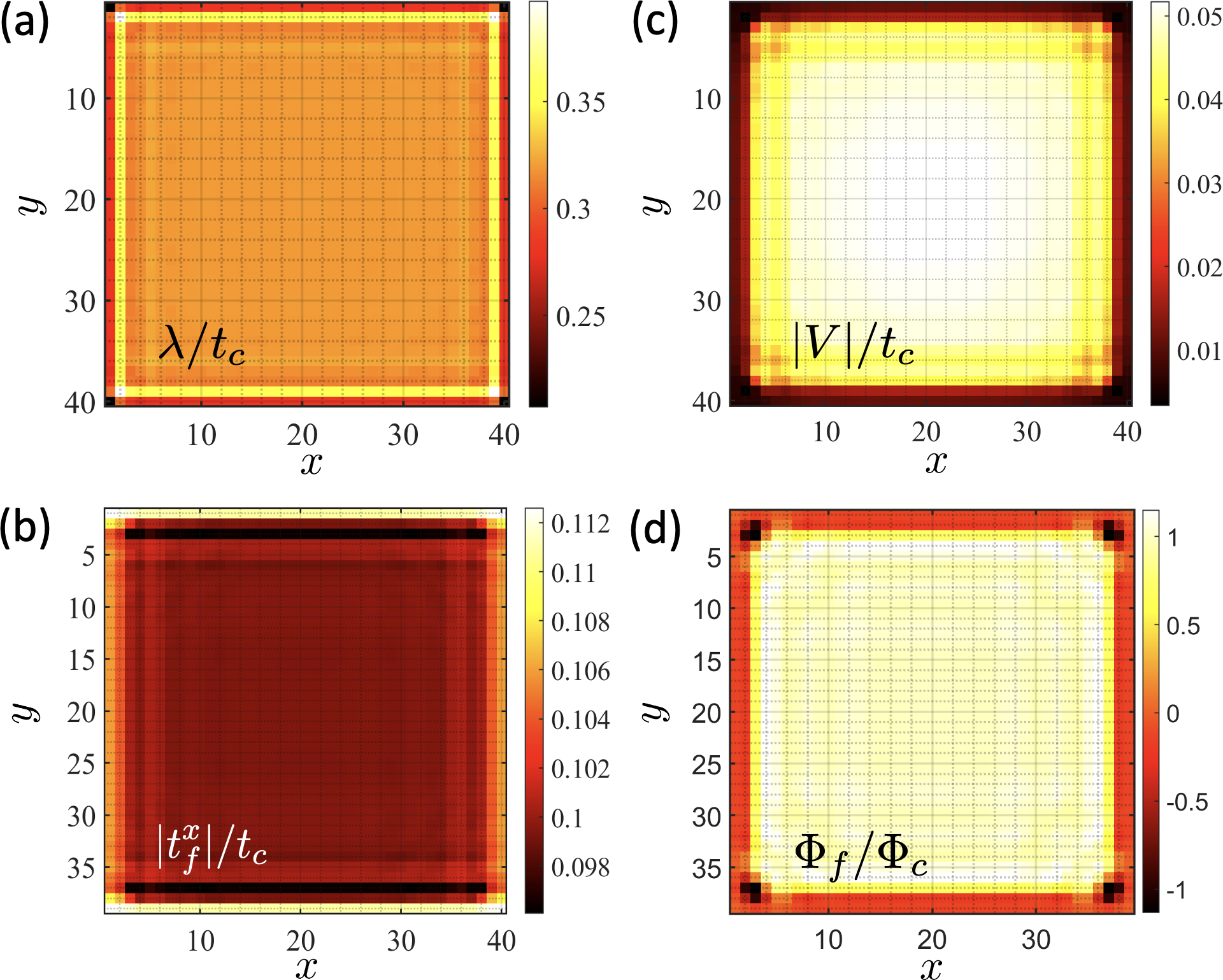}
	\caption{\label{fig2}The 2D tight-binding results on MF parameters of a TKI on a 40$\times$40 square lattice with open boundary condition and parameters $\mu/t_c=3.2$,  $Q=0.14$,  $J_K/t_c=0.23$, and $J_H/t_c=0.83$ at $T/t_c=0.01$ under a magnetic flux $Ba^2=1/8$ applied in the symmetric gauge: (a) $\lambda/t_c$, (b) $f$-electron hopping amplitudes along horizontal links, $\abs{t_f^x}/t_c$, (c) hybridization $\abs{V/t_c}$ where the solution $V$ is real, and (d) The induced flux through $f$-plaquettes $\Phi_f$, relative to the external magnetic flux $\Phi_c$.}
\end{figure}

Nonetheless, our numerical results on a finite lattice with open boundary condition shows that at least in the low magnetic field regime, where the Kondo physics is relevant and magnetic oscillations are visible, the MF parameters are uniform under a weak magnetic field. In Fig.\,\ref{fig2}, we show the spatial profile of (a) $\lambda$, (b) $t_f$, and (c) $V$ in a TKI. All their magnitudes are uniform in the bulk. In addition, we can see the effect of gauge locking in Fig.\,\ref{fig2}(d). The internal gauge field flux through a unit cell, computed from product of hopping amplitudes counter-clockwise around the unit cell, $\Phi^\mathbf{r}_f\equiv \mathop{\mathrm{Im}} \ln(\prod_{\vv{ij}\in\partial\Box_\mathbf{r}} t^{ij}_f)$,
is equal to that set by the external magnetic field, i.e., $\Phi_c^{\mathbf{r}}$. Thus, driven by the Kondo interaction, $t_f$ acquires the same hopping phase as $t_c$. Near lattice boundaries, the Kondo hybridization vanishes due to smaller coordination number and the gauges are no longer locked.

The fact that MF parameters are homogenous in the bulk implies that the energy spectrum in Eq.\,\eqref{eq:M} is essentially indistinguishable from that of a normal or topological insulator, albeit with temperature and magnetic field dependent MF parameters. Thus, we can use the cylindrical geometry with Laudau gauge for our numerical investigations of magnetic oscillation in the following.

\subsection{Magnetic oscillation and Kondo breakdown}

\subsubsection{Cylindrical geometry setup}

As demonstrated in Sec.\,\ref{sec:homo}, the MF parameters are homogeneous inside the bulk in the presence of magnetic field. It is then convenient to impose the periodic boundary condition along one direction of the two dimension Kondo lattice, which leads to a cylindrical geometry [Fig.\,\ref{fig1}(b)].

In cylindrical geometry, Landau gauge is used so that $A_c = B [-(y-y_0),0,0]$, where $y_0$ is half the system length along $y$ direction. Translational invariance along $x$-direction admits the following Fourier transformation.
\begin{equation}
\hspace{-.6em}	c_{k,n} = \frac{1}{\sqrt{L}}\sum_{m} e^{ikm} c_{m,n}, \quad c_{m,n} = \frac{1}{\sqrt{L}} \sum_{k} e^{-ikn} c_{k,n},
\end{equation}
where $m,n=1,\dots,L$ are the site indices along $x,y$ direction respectively.
The kinetic term for the conduction electron on a cylinder becomes
\begin{equation}
	H_c^t = - t_c \sum_{k}\sum_{n=2}^L c_{kn}^\dagger \left(e^{i[k+eB(y_n-y_0)]}c_{k,n}+c_{k,n-1}\right) + \mathrm{h.c.}
\end{equation}
The MF equations are modified accordingly in Eq.\,\eqref{eq:CylinderMF}.

\subsubsection{Normal Kondo Insulator}

\begin{figure}[t]
\includegraphics[width=\linewidth]{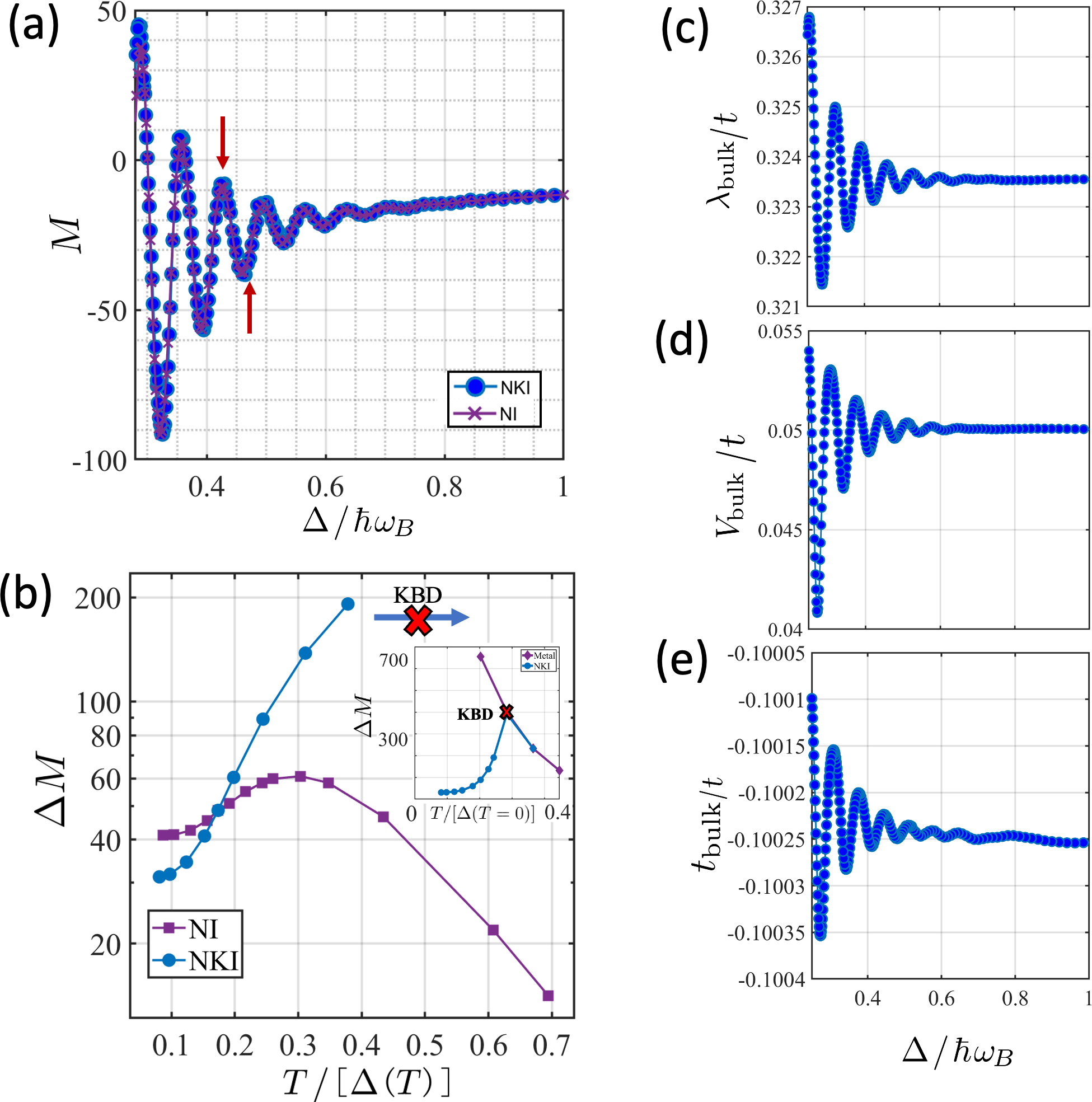}
\caption{\label{fig3}
Magnetic oscillations in NKI vs.\ NI for a 100$\times$100 square lattice in cylinder 
geometry and parameters  $m_f/m_c=-10$, $\mu/t_c=3.2$,  $Q=0.14$, $J_K/t_c=1.39$ and $J_H/t_c=1.68$. 
  (a) Magnetization as a function of inverse magnetic field $\Delta/\hbar\omega_B$ where $\Delta=0.058t_c$ is the (indirect) spectral gap at zero magnetic field and $\omega_B\equiv eB/m_c$. Here, $T/t_c=0.01$ corresponding to the line crossing point in (b).
  (b) The oscillation amplitude of magnetization in the half period indicated by arrows in (a) as a function of $T/\Delta(T)$, where $\Delta(T)$ is the zero field gap at different temperatures. For NI $\Delta=0.058t_c$ is $T$-independent.
  The inset shows the temperature dependence of the oscillation amplitude difference for NKI and a metallic $c$-electron case. They agree beyond the Kondo breakdown (KBD). The $T$ scale used here is the gap at zero field and zero temperature, $\Delta(T=0) = 0.062t_c$.
  (c)--(e) Mean-field parameters of NKI at the center of cylinder: (c) $\lambda_{\rm Bulk}/t_c$ , (d) $t_{\rm Bulk}$, and (e)  $V_{\rm Bulk}$. 
  }
\end{figure}

The magnetic oscillation of an NKI is shown in Fig.\,\ref{fig3}(a). The data shows a clear $1/B$-periodic quantum oscillation. It behaves nearly the same as that of a normal insulator (NI) with fixed $V$. This can be expected from Eq.~(\ref{eq:M}), owing to the saddle point equation and the homogenity condition. By keeping the $B$-field fixed, we can monitor the change of oscillation amplitude as a function of temperature. Above a critical temperature, Kondo break down occurs where the hybridization $V$ goes to zero, and so does the zero field gap $\Delta(T)$. This marks the difference between the NKI and NI since the zero field gap of an NI does not change with temperature. 


In Fig.\,\ref{fig3}(b), we study the scaled temperature dependence of the oscillation amplitude between magnetic fields marked by the arrows in Fig.\,\ref{fig3}(a). The $x$-axis is the temperature divided by the zero-field gap $\Delta(T)$, which is \emph{temperature dependent} in NKI. The oscillation in NI and NKI becomes of equal amplitude at around $T/\Delta(T)=0.2$. This is the temperature used in Fig.\,\ref{fig3}(a). The temperature dependence of the magnetization oscillation in NI obeys the Lifshitz-Kosevich (LK) theory for a narrow gap insulator developed in previous section (see also \cite{zhang2016quantum}). At high temperature, the oscillation amplitude decays exponentially with temperature. However, in the NKI the temperature dependence clearly shows deviation from the traditional LK theory. Increasing temperature leads to a decreasing Kondo hybridization and the Kondo gap, resulting in a different temperature scaling. At a critical temperature of $T/\Delta(T)=0.8$, KBD occurs and the diminishing gap $\Delta(T)$ leads to a drastic increase of magnetic oscillation amplitude of NKI. Above the breakdown temperature, 
the behavior of the magnetization follows that of a normal metal. We show this in the inset of Fig.\,\ref{fig3}(b) with $T/\Delta(T=0)$ as the $x$-axis, since at higher temperatures $\Delta(T)=0$. It confirms that beyond the critical temperature, NKI magnetization exactly follows the temperature dependence of a normal metal.

The oscillations of MF parameters are show in Figs.\,\ref{fig3}(c)--(e), at the same temperature as in Fig.\,\ref{fig3}(a). Distinct from the NI case, they show the same $1/B$-oscillation as the magnetization. At this temperature, far from the KBD, these oscillations are small. Hence, the single-particle orbitals are not very much affected, explaining the similarity between magnetic oscillations of NKI and NI in Fig.\,\ref{fig3}(a).

\subsubsection{Topological Kondo Insulator}

We now turn to the field and temperature dependence of magnetization in a TKI. The results here are qualitatively similar to that of an NKI. However, at system sizes smaller than the thermal coherence length $l_T\sim v_F/k_BT$, the Aharonov-Bohm oscillation emerging from the edge states can overwhelm the bulk numerical signal (Appendix \ref{ss:AB}). We use a large system size and finite $T$ to suppress this interference.

\begin{figure}
\includegraphics[width=\linewidth]{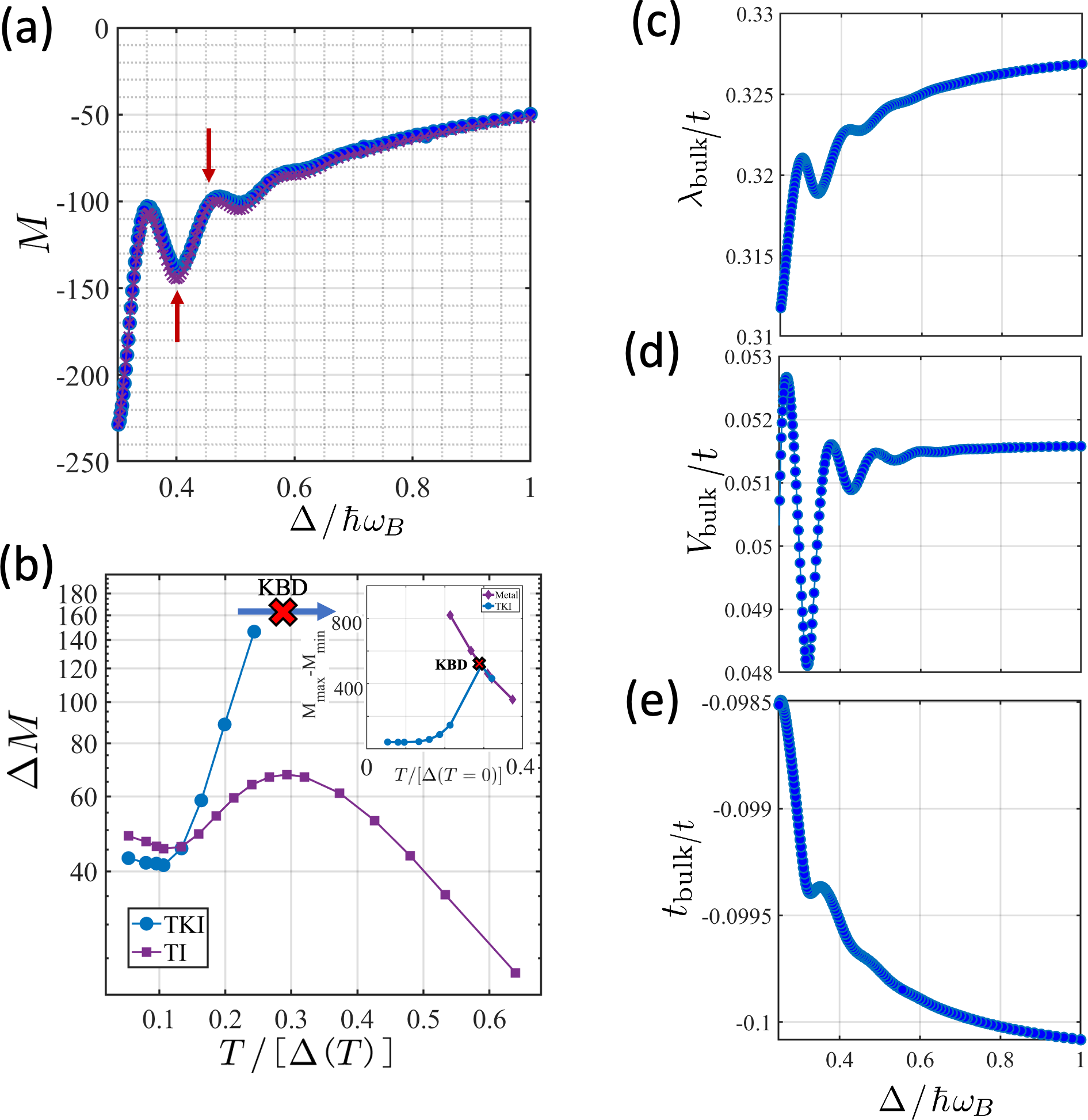}
\caption{\label{fig4}
Magnetic oscillations in TKI vs.\ TI for an 80$\times$80 square lattice in cylindrical geometry and parameters $m_f/m_c=-10$, $\mu/t_c=3.2$,  $Q=0.14$, $J_K/t_c=0.23$ and $J_H/t_c=0.83$.
(a) Magnetization as a function of inverse magnetic field $\Delta/\hbar\omega_B$ where $\Delta=0.094t_c$ is the indirect (smallest) gap in the spectrum and $\omega_B\equiv eB/m_c$. Here, $T/t_c=0.01$ corresponding to the line crossing in (b).
(b) The oscillation amplitude of magnetization in the half period indicated by arrows in (a) as a function of $T/\Delta$, where $\Delta$ is the zero field gap at different temperatures. For TI, $\Delta=0.094t_c$ is $T$-independent. The inset shows the temperature dependence the oscillation amplitude for TKI and a metallic $c$-electron case. They agree beyond the Kondo breakdown (KBD). The $T$ scale used here is the gap at zero field and zero temperature, $\Delta(T=0) = 0.094t_c$.
(c)--(e) Mean-field parameters of TKI at the center of cylinder: (c) $\lambda_{\rm Bulk}/t_c$, (d) $t_{\rm Bulk}$, and (e)  $V_{\rm Bulk}$. }
\end{figure}

Figure \ref{fig4}(a) shows the magnetization of the TKI and a topological insulator for a fixed $V$. The data also exhibits a $1/B$-periodic quantum oscillation. The oscillation pattern is nearly identical between TKI and TI. Furthermore, for this particular temperature, the amplitudes are also the same.

The scaled temperature dependence of magnetization oscillation is studied in Fig.\,\ref{fig4}(b). We extract the amplitude between magnetic fields indicated by the two arrows in Fig.\,\ref{fig4}(a). While the $T$-dependence of TI also follows LK theory, deviation again shows up in TKI since the gap is closing with rising temperature. Similar to the NKI case, the zero field gap of TKI diminishes with increasing temperature until the KBD. This leads to an increase in quantum oscillation magnitude in TKI compared to TI. The inset of Fig.\,\ref{fig4}(b) shows the metallic behavior beyond the KBD.

Finally, the oscillations of MF parameters are shown in Figs.\,\ref{fig4}(c)--(e). Similar to the case of an NKI, their changes are small at low temperatures.

\section{Conclusion}
In this paper we have investigated quantum magnetic oscillations in two-dimensional normal and topological Kondo insulators using a mean-field theory that becomes exact in the large-$N$ limit. In 2D, the TKI model decouples into two spin sectors, each with their own hybridized electron-spinon edge modes, that only interact energetically. We have shown that in a TKI an effective Higgs term locks the internal gauge field to the external one and the spinons become charged, similar to the case of the NKI. The mean-field manifestation of this effect is a Meisner effect for the gauge difference which leads to the Kondo flux repulsion out of Kondo plaquettes. Therefore, the magnetic oscillations in the NKI and TKI can be compared with magnetic oscillations in normal and topological band insulators, respectively. 

Unlike metals, in an insulator the Landau levels do not cross the chemical potential. Low temperature quantum oscillations rather originate from the occupied LLs in the valence band, first approaching and then withdrawing from the top of the valence band \cite{zhang2016quantum}. We have taken these into account, providing a closed form formula for the oscillatory part of the free energy which can be regarded as the modified Lifshitz-Kosevitch formula for NI and TIs. The magnetic oscillations are damped exponentially, $\sim \frac{\Delta}{\hbar\omega_B}e^{-\Delta/\hbar\omega_B}$, for magnetic fields smaller than the gap $\hbar\omega_B\ll \Delta$.  To the best of our knowledge both the derivation of the Higgs term for the TKI and the modified LK formula for NI and TI are presented here for the first time.

Our analysis is supplemented with a numerical solution of mean-field equations in presence of magnetic field. In the case of TKI,  this requires special care to suppress the $B$-periodic Aharonov-Bohm oscillations originating from the edge states.  At low fields, we find that mean-field parameters of NKI and TKI remain homogeneous and therefore, a comparison between NI and TI is relevant. However, these mean-field parameters vary with temperature, resulting in a gap $\Delta(T)$ that vanishes with increasing temperature. After establishing the homogeneity of mean-field parameters, we analyze the oscillations in a cylindrical geometry with translational symmetry assumed in the direction of periodicity. 

We find that the amplitude of the oscillations in NKI and TKI is not damped but rather enhanced with temperature, in stark contrast to those of the NI and TI. These result can be summarized as oscillations that are exponentially damped in $\Delta/\hbar\omega_B$. But in contrast to band insulators, they are enhanced with increasing the temperature. At large field or high temperature, these trends are invariably interrupted by the Kondo breakdown due to the suppression of the density of states.

In experiments on 3D materials, KBD is likely inhibited due to the larger density of states. Furthermore, surface states do not interfere with magnetic oscillations due to {the larger bulk-surface ratio} and smaller coherence length. 
Nonetheless, several aspect of this study is relevant to the existing experiments. {A $T \!\to\!0$ magnetic oscillation that is not exponentially suppressed in the $B\to 0$ limit clearly cannot be explained within a noninteracting or mean-field framework.} Admittedly, this is a difficult test as in real systems the oscillations are weakened by the disorder in this limit.

There are many ways in which this work can be extended, e.g.\ by studying the three-dimensional case, inclusion of disorder and calculation of resistivity.  Since many of the properties of the Kondo insulators parallel that of a superconductor, it is natural to expect a type-II Kondo insulator with localized KBD regions, in which the locking between internal and external gauge fields is locally broken, allowing a vorticity in the gauge difference. Recently, this mechanism has been suggested as an explanation for the intricacies of 4Hb-TaS$_2$ phase diagram \cite{Koenig2023}. Indeed, we observe that close to a KBD, inhomogeneities set in the mean-field parameters, a careful study of which is left for the future. 

\begin{acknowledgments}
It is a pleasure to thank Chen-Te Ma for valuable discussions. This research was supported in part through research cyberinfrastructure resources and services provided by the Advanced Research Computing center at the University of Cincinnati, Cincinnati, OH, USA.
\end{acknowledgments}

\appendix

\section{Numerical methods for the self-consistent mean-field solution}
\label{Solver}

\begin{figure}
	\centering 
\begin{tikzpicture}[thick,scale=0.55, every node/.style={scale=0.55}]
	\node[draw, align=center,rounded corners]        (input)  at (0,0) {Input parameters \\
	$(m_c,m_f,Q,J_K,J_H,T,B)$ \\
	Input initial MF parameters\\
	$\theta_0\equiv(\lambda_0,V_0,t_{f0})$};
	\node[draw, diamond,aspect=2, below=70pt of input]           
	(qf)  {$\sum_\mathbf{r} \left|\dfrac{\hat{Q}(\mathbf{r})}{Q}-1 \right| <\epsilon_q$?};
	 \node[draw, diamond,aspect=2, below=20pt of qf]          
	(Vt) 
	{
	  $
	  V_\mathrm{Loss} + t^f_\mathrm{\mathrm{Loss}} < \epsilon? 
	  $
	};
	\node[draw,below=30pt of input,align=center]
	(ED) {
		Diagonalize the MF Hamiltonian \\
	$H(\theta)\Rightarrow (\Psi_l(\mathbf{r}), \epsilon_l)$};
  	\node[draw, left=20pt of Vt,align=center]           
	(Vt-loop) 
	{
	  $V_\mathbf{r} = (1-\eta') V_\mathbf{r} - \eta' J_K \langle f^\dagger_\mathbf{r}c_\mathbf{r}\rangle $
	  \\
	  $t_f^{\mathbf{r}\mathbf{r}'} = (1-\eta') t_f^{\mathbf{r}\mathbf{r}'} - \eta' J_H \langle f^\dagger_\mathbf{r}f_{\mathbf{r}'}\rangle $
	};

	\node[draw, below=20pt of Vt,align = center,rounded corners] (F)
	{
	  Compute the free energy \\
	  $F(T,B)$
	};
	\node[draw, right=20pt of qf] (lam-loop) {$\lambda_\mathbf{r} = \lambda_\mathbf{r} - \eta \left(\dfrac{\hat{Q}(\mathbf{r})}{Q}-1\right)$};
	
	\draw[-{Latex[length=1.5mm]}] (input)  -- (ED);
	\draw[-{Latex[length=1.5mm]}] (ED)  -- node[left]{} (qf);
	\draw[-{Latex[length=1.5mm]}] (qf)  -- node[above]{No}  (lam-loop);
	\draw (input) -- coordinate (iED) (ED);
	
	\draw[-{Latex[length=1.5mm]}] (lam-loop) |- (iED);
	\draw[-{Latex[length=1.5mm]}] (qf)  -- node[left]{Yes}  (Vt);
	\draw[-{Latex[length=1.5mm]}] (Vt)  -- node[above]{No}  (Vt-loop);
	\draw (qf) -- coordinate (qfVt) (Vt);
	\draw[-{Latex[length=1.5mm]}] (Vt-loop) |- (iED);
	\draw[-{Latex[length=1.5mm]}] (Vt) -- node[left]{Yes} (F);
  \end{tikzpicture}
  \caption{\label{s1}Self-consistent solver workflow. $\hat{Q}(\mathbf{r})$ is the $f$-electron occupation calculated from the current mean-field parameters. $\eta$ and $\eta'$ are the dampings or learning rates set manually. $V_{\text{loss}}$ and $t^f_{\text{loss}}$ are defined in Eq.\,\eqref{eq:loss}.} 
\end{figure}
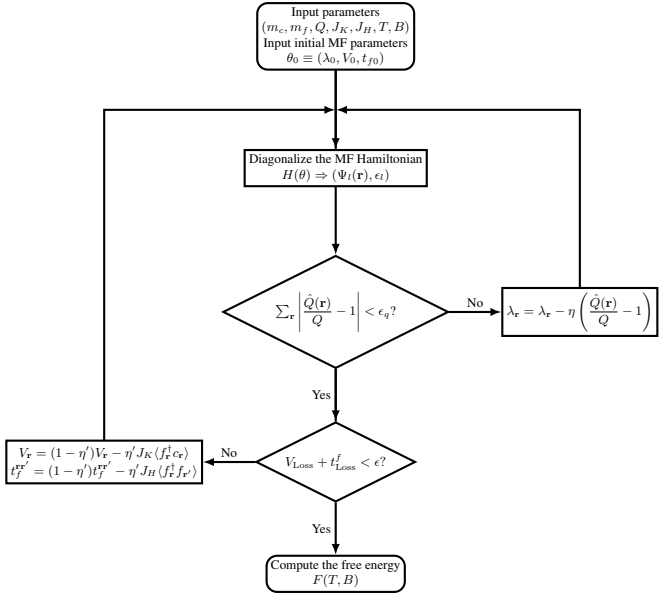

This section describes the mean-field solver. The self-consistent solver workflow is depicted in Fig.\,\ref{s1}. 
We start from a set of initial MF parameters $(\lambda_0, V_0, t_{f0})$ for the Hamiltonian, each a spatial function with optional randomness to test the robustness of the solver. Then, the Hamiltonian is diagonalized exactly. Next, we search for $\lambda_\mathbf{r}$ to maintain the desired $f$-electron occupation corresponding to the spin size. With the spin size constraint satisfied, we update $V_\mathbf{r}$ and $t_f^{\mathbf{r}\mathbf{r}'}$ according to the mean field equations~\eqref{eq:MF}, with some damping to help convergence. The solver then continues to the next iteration of diagonalization, $\lambda$-search, and $(V, t_f)$-update.

In each epoch, we monitor the global loss defined by
\begin{eqnarray}
	G_\mathrm{loss} = V_\mathrm{loss} + t^f_\mathrm{loss},
\end{eqnarray}
where 
\begin{eqnarray}
	\label{eq:loss}
     V_\mathrm{loss} &=& \sum_{\mathbf{r}} 
	|V_\mathbf{r} + J_K \langle f^\dagger_\mathbf{r} \tilde{c}\dn_{\mathbf{r}} \rangle|
	 \\
	&=& \sum_{\mathbf{r}} \left| V_\mathbf{r} + J_K \sum_l [\psi_l^f(\mathbf{r})]^* \tilde{\psi}_l^c(\mathbf{r}) f(\varepsilon_l)\right|,\nonumber\\
	t^f_\mathrm{loss} &=& \sum_{\substack{\mathbf{r}\\
	\boldsymbol{\delta} = \hat{\mathbf{x}},\hat{\mathbf{y}} }} \left| t_f^{\mathbf{r},\mathbf{r}+\boldsymbol{\delta}} + J_H \langle f_\mathbf{r}^\dagger f\dn_{\mathbf{r}+\boldsymbol{\delta}}\rangle\right|\nonumber\\
	&=&\sum_{\substack{\mathbf{r} \\
	\boldsymbol{\delta} = \hat{\mathbf{x}},\hat{\mathbf{y}} }} \left| t_f^{\mathbf{r},\mathbf{r}+\boldsymbol{\delta}} + J_H \sum_l [\psi_l^f(\mathbf{r})]^* \psi_l^c(\mathbf{r}+\boldsymbol{\delta}) f(\varepsilon_l)\right| . \nonumber
\end{eqnarray}

Here, $l$ is the single-particle orbital index, $f(\varepsilon)$ is the Fermi-Dirac function, and $\tilde{c}$ is a short-hand notation for $c$ in an NKI, and $\breve{c}$ in a TKI. When the global loss becomes less than a preset threshold, we declare convergence and output all the parameters as well as the free energy computed.

\begin{figure}
	\includegraphics[width=\linewidth]{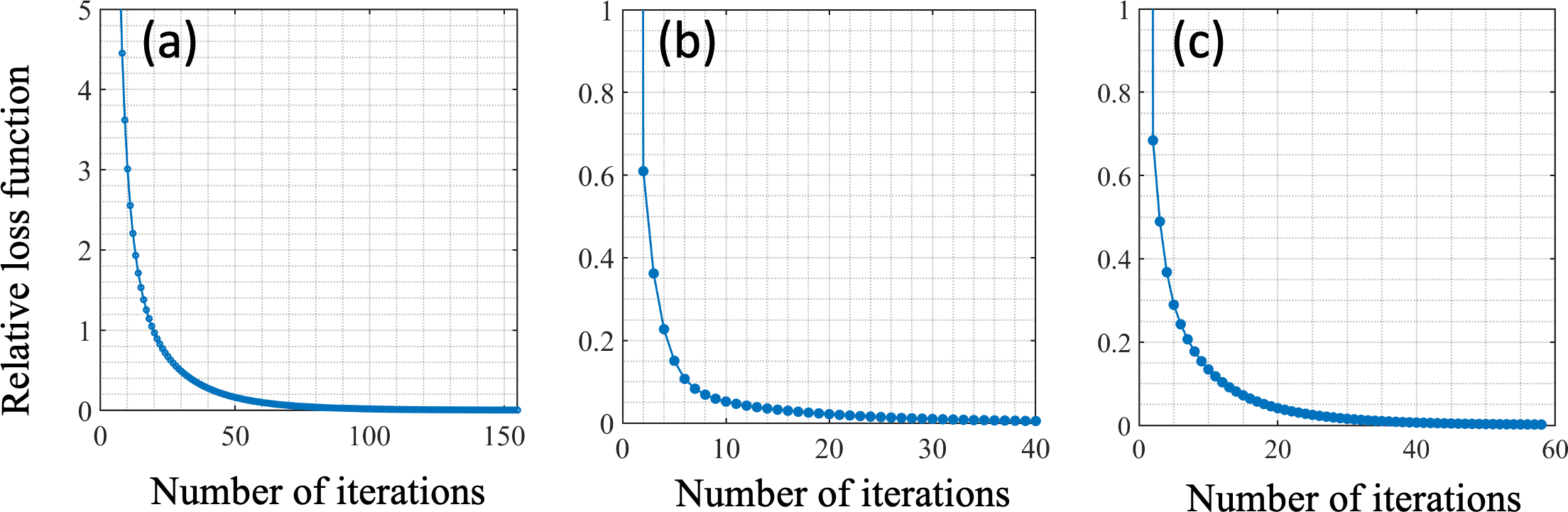}
	\caption{\label{s3} Convergence studies of the self-consistent solver. (a) 2D TKI case in Fig.\,\ref{fig2}. (b) NKI with cylindrical geometry in Fig.\,\ref{fig3} at $\hbar \omega_B / t_c = 0.06$ (c) TKI with cylindrical geometry in Fig.\,\ref{fig4} at $\hbar \omega_B / t_c = 0.1$}
\end{figure}

\begin{figure}
	\includegraphics[width=\linewidth]{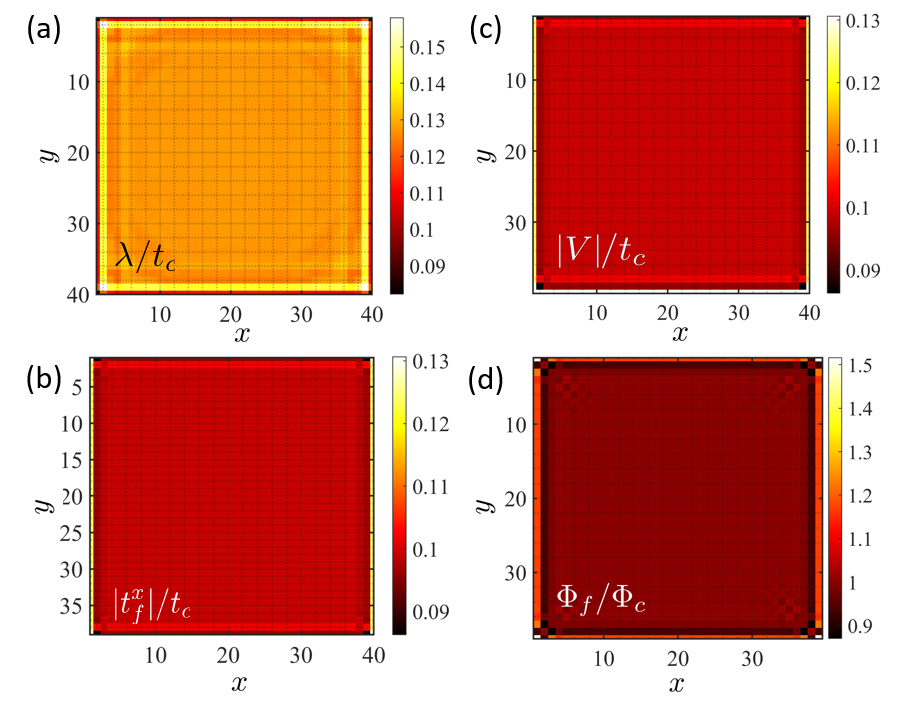}
	\caption{\label{s2}2D tight-binding results on mean-field parameters of NKI on a 40$\times$40 square lattice with open boundary condition and the parameters $\mu/t_c=0.4$,  $Q = 0.42$,  $J_K/t_c=1.56$ and $J_H/t_c=0.55$ at $T/t_c=0.002$ under a magnetic flux $Ba^2=0.011$ in the symmetric gauge: (a) $\lambda/t_c$, (b) $f$-electron hopping amplitudes along horizontal links, $\abs{t_f^x}/t_c$, (c) hybridization $\abs{V/t_c}$ where the solution $V$ is real, and (d) The induced flux through $f$-plaquette $\Phi_f$, divided by the external magnetic flux $\Phi_c$.}
\end{figure}

The self-consistent solver usually converges with a diminishing loss after a number of iterations, as plotted in Fig.\,\ref{s3}. In an open-boundary 2D square lattice, this results in a uniform landscape of MF parameters, as was shown in Fig.\,\ref{fig2} for a TKI. The results is similar in the NKI case, shown here in Fig.\,\ref{s2}.

Proper gauge choices speed up the convergence. For an open square lattice, a symmetric gauge for $A_c$ is used such that it respects the $C_4$ symmetry on the defined lattice so as to improve convergence. On a cylinder, the Landau gauge is also chosen to respect the mirror symmetry about $y_0$ at halfway the system height.

Once we established the uniformity of mean field parameters, all our calculations can be done on a cylinder. Here, parameters are assumed to be homogeneous along the periodic direction $x$. Thus, the MF equations \eqref{eq:MF} and the loss functions \eqref{eq:loss} are modified accordingly. All mean field parameters depend only on $y$. In particular,
\begin{eqnarray}
	\label{eq:CylinderMF}
	V_y & = & -\frac{J_K}{L} \sum_{k} \langle  f\dg_{y k} \tilde{c}\dn_{y k} \rangle , \nonumber \\
	t^{\mathbf{r},\mathbf{r}+\hat{\mathbf{x}}}_f & = & -\frac{J_H}{L} \sum_{k} e^{ik}\langle f\dg_{y k} f\dn_{y k} \rangle , \nonumber \\ 
		t^{\mathbf{r},\mathbf{r}+\hat{\mathbf{y}}}_f & = & -\frac{J_H}{L} \sum_{k} \langle f\dg_{y k} f\dn_{y+\hat{y}, k} \rangle .
\end{eqnarray}

\section{Aharonov-Bohm oscillations \label{ss:AB}}

Aharonov-Bohm (AB) oscillations are present in topological insulators as a quantum interference phenomena \cite{bardarson2013quantum}.
In the numerical calculation, we found that there is a threshold in system size to observing a clear bulk quantum oscillation. This is because at smaller system sizes, AB oscillation from edge modes are prominent and result in multicomponent oscillation patterns.

\begin{figure}
	\includegraphics[width=0.5\linewidth]{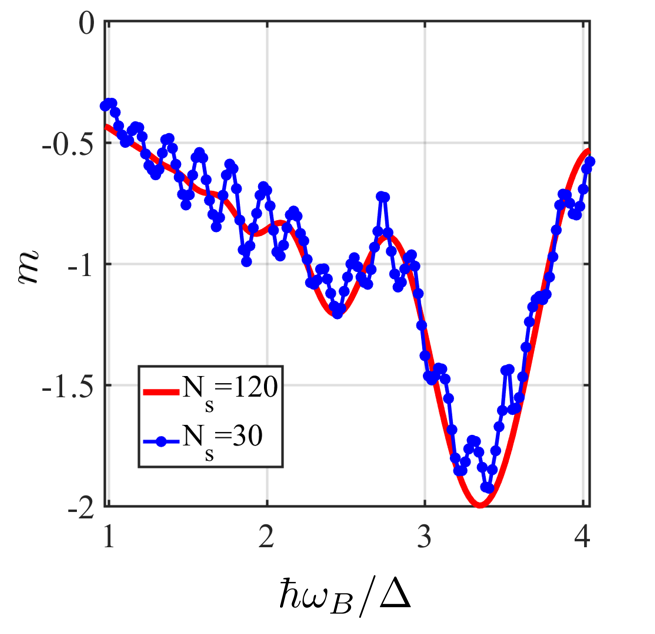}
	\caption{\label{s5}The oscillation of magnetization $m$ for a TI with different system sizes. Parameters used are the same as those in Fig.\,\ref{fig4}. The $B$-periodic oscillations in smaller system are due to Aharonov-Bohm oscillations coming from the edge modes.}
\end{figure}

In cylindrical geometry we choose systems with linear dimensions $L = 30$ and $120$ respectively and calculate the magnetization density. As shown in Fig.\,\ref{s5}, for $L=30$ case, it shows clear $B$-periodic oscillations at low field. This originates from the strong interference effect of the edge mode at the boundary when system is too small. The AB oscillation relies on the coherence length to be on the scale of the system perimeter. This can be limited by the thermal coherence length at increased temperatures \cite{Roulleau2008}.

\section{Derivation of TKI effective action}
\label{SE}

Here we derive the TKI effective action expanded near $\Gamma$, in one of the two TR-related spin sectors. Those at $M$, $X$, and $Y$ proceeds similarly. The calculation follows closely that of \cite{Wugalter20}. We start from the Lagrangian \eqref{eq:L}. Integrating out the fermions, the free energy reads 
\begin{equation}
	F(A) = -NT\, \mathrm{Tr} \log[-\mathcal{G}^{-1}(A)] +  i  A_f^\tau Q + i A_c^\tau n_c,
	\label{eq:FE}
\end{equation}
and in the $(c, f)^T$ basis,
\begin{eqnarray}
	\label{eq:GTKI}
	-\mathcal{G}^{-1}(A)= 
	\begin{pmatrix}
		- g_c^{-1}(A_c) & \mathcal{V}(\mathbf{A}_f) \\
		\mathcal{V}^\dagger(\mathbf{A}_c) & -g_f^{-1}(A_f)
	\end{pmatrix}.
\end{eqnarray}
where $g_c$ and $g_f$ are the free propagators for $c$ and $f$ electrons minimally coupled to the $U(1)$ gauge fields,%
\begin{eqnarray}
	-g_c^{-1}(A_c)&=&\left(\partial_\tau-\mu\right)+i A_c^\tau + 
	\frac{(\nabla+i \mathbf{A}_c)^2}{2 m_c},
	\nonumber \\
	-g_f^{-1}(A_f)&=&
	\left(\partial_\tau+\lambda\right)+i A_f^\tau
	+ \frac{(\nabla+i \mathbf{A}_f)^2}{2 m_f}.
\end{eqnarray}%
The hybridization $\mathcal{V}$ takes different forms for NKI and TKI.
\begin{eqnarray}
	\text{NKI:} \quad \mathcal{V}(\mathbf{A})  &=&  V ,   \\
	\label{eq:VTKI}\text{TKI:} \quad \mathcal{V}(\mathbf{A})  &=&  V [- i(\partial_x + i \partial_y) - (A_x + i A_y)] \nonumber\\
	&\equiv& V(-i\partial^+ - A^+).
\end{eqnarray}
Here we define a general notation
\begin{eqnarray}
	 \partial^\pm &\equiv& \partial_x \pm i \partial_y, \nonumber\\
	 A^\pm &\equiv& A_x \pm i A_y, \nonumber\\
	 k^\pm &\equiv& k_x \pm i k_y.
\end{eqnarray}

Despite its appearance, the hybridization terms $\mathcal{V}(\mathbf{A}_c)$ and $\mathcal{V}\dg(\mathbf{A}_f)$ in Eq.\,\eqref{eq:VTKI} are hermitian conjugates up to a total derivative and gauge transformation of $V$, which we take a detour to discuss below. The vector potentials can be Helmholtz-decomposed into their longitudinal, i.e.\ curl-free, and transverse components, $\mathbf{A}=\mathbf{A}^L+\mathbf{A}^T$ \cite{jackson1999}. To show the hermicity of the longitudinal contribution, we temporarily adopt the following gauge for $V$
\begin{eqnarray}
	V_\mathbf{r} & = & |V| e^{i \varphi_{cf}(\mathbf{r})}, \nonumber\\
	\varphi_{cf}(\mathbf{r}) & = & \int^\mathbf{r} \mathrm{d} \mathbf{s} \cdot [\mathbf{A}^L_c (\mathbf{s}) - \mathbf{A}^L_f  (\mathbf{s})].
\end{eqnarray}
The phase factor $\varphi_{cf}$ thus captures the longitudinal part of the vector potentials. Then,
 $V_\mathbf{r}\bar{f}_\mathbf{r}(-i\partial^+-\mathbf{A}^{L+}_c)c_\mathbf{r}$ becomes hermitian conjugate to
 $V^*_\mathbf{r}\bar{c}_\mathbf{r} (-i\partial^--\mathbf{A}^{L-}_f)f_\mathbf{r}$ up to a total derivative. As for the transverse contributions, they assume the form
\begin{equation}
	\mathbf{A}^T(\mathbf{r}) = \int \frac{d^2\mathbf{r}'}{2\pi} B(\mathbf{r}') \frac{\hat{\mathbf{z}}\times (\mathbf{r}-\mathbf{r}')}{|\mathbf{r}-\mathbf{r}'|^2},
\end{equation}
for some (pseudo)magnetic field $B=\nabla\times\mathbf{A}$.
Thus, its contribution to $\mathcal{V}(\mathbf{A}_c)$ in $\mathcal{L}$ is
 \begin{equation}
 	\frac{V_{\mathbf{r}}}{2 \pi}  \bar{f}_{\mathbf{r}}  
 	\left\{ \int \frac{\mathrm{d}\mathbf{r}' B (\mathbf{r}')}{| \mathbf{r} - \mathbf{r}' |^2} [(y + ix) - (y' +ix')] \right\}
 	c_{\mathbf{r}},
 \end{equation}
while the contribution to $\mathcal{V}\dg(\mathbf{A}_f)$ is similar. With a gauge transformation on $c$ or $f$, the factors $y\pm ix$ can be rotated to $r$, making them also curl-free. Therefore, the TKI hybridization terms are hermitian.

In the following discussion, we adopt the radial gauge \cite{Coleman2015} so that  $V$ is real and its phase is absorbed into $f$,
\begin{equation}
	\label{eq:Vreal}
	V e^{i\varphi} \to V, \quad A^\mu_f \to A^\mu_f - \partial^\mu \varphi.
\end{equation}

After integrating out the fermionic degrees of freedom, effective free energy becomes  
\begin{eqnarray}
	\Delta F (A) &\equiv& F(A) - F(A=0) \nonumber \\
	&=& - NT \, \tr {\ln(1 - \mathcal{G}\mathcal{W})} +  F_b ,
\end{eqnarray}
where $F_b = i (A_c^\tau n_c + A_f^\tau Q)$, $-\mathcal{G}^{-1} \equiv -\mathcal{G}^{-1}(A=0)$ and $\mathcal{W}$ takes the form  
\begin{eqnarray}
	\mathcal{G}
	= 
	\begin{pmatrix}
		G_{c c} & G_{c f} \\ 
		G_{f c} & G_{f f}
	\end{pmatrix}, \quad
	\mathcal{W} 
	= \begin{pmatrix}
		W_{c c} & W_{c f} \\ 
		W_{f c} & W_{f f}
	\end{pmatrix}.
\end{eqnarray}
In the momentum-frequency domain, the gauge field-independent Green's functions are given by
\[ G_{c c} =  (g_c^{-1} g_f^{-1} - V^2k^2)^{-1} g_f^{-1}, \]
\[ G_{f f} = (g_c^{-1} g_f^{-1} - V^2 k^2)^{-1} g_c^{-1}, \]
\be
	G_{cf} = [G_{fc}]^* = V k^+ g_f G_{c c},
\ee
where $g \equiv g(A=0)$ and the gauge field-dependent part $\mathcal{W}$ is
\begin{eqnarray}
	W_{c c} &=&  iA_c^\tau - \mathbf{v}_c \cdot \mathbf{A}_c + \frac{\mathbf{A}_c^2}{2 m_c}, \nonumber \\ 
	W_{f f} &=& iA_f^\tau - \mathbf{v}_f \cdot \mathbf{A}_f + \frac{\mathbf{A}_f^2}{2 m_f}, \nonumber \\ 
	W_{c f} &=& -V A_f^+,\: W_{f c} = -V A_c^-.
\end{eqnarray}
where $\mathbf{v}\equiv \mathbf{k}/m$ are the group velocities.

It is convenient to separate the paramagnetic and diamagnetic contributions in $\mathcal{W}$,
\begin{eqnarray}
	\mathcal{W} &=& \mathcal{W}^{p} +  \mathcal{W}^{d},
	\\
	\mathcal{W}^{d} &=& \mathrm{diag} \biggl(
		\frac{\mathbf{A}_c^2}{2 m_c}, \,\frac{\mathbf{A}_f^2}{2 m_f}	
	\biggr).
\end{eqnarray}

We expand the logarithm in \eqref{eq:FE} by defining $\mathcal{X} \equiv \mathcal{G}\mathcal{W}$ and proceed with  
\begin{eqnarray}
	- \ln (1 - \mathcal{X}) \approx \mathcal{X} + \mathcal{X}^2 /2,
\end{eqnarray}
To quadratic order in the gauge fields, we get 

\bw

\begin{eqnarray}
	\frac{\Delta F}{N T} = F_b +\Tr \bigg[
	\frac{1}{2} (G_{c c} W^p_{c c})^2 &+& 
	G_{c c} G_{c f} W^p_{c c} W^p_{f c} + G_{c c} W_{c c}^d \nonumber \\ 
	+\frac{1}{2} (G_{f f} W^p_{f f})^2 &+& G_{f f} G_{f c} W_{f f}^p W^p_{c f} + G_{f f} W_{f f}^d \nonumber \\ 
	G_{c c} G_{f c} W_{c c}^p W_{c f}^p &+& G_{f f} G_{c f} W_{f f}^p W_{f c}^p +G_{c c} G_{f f} W_{c f}^p W_{f c}^p + G_{c f} G_{f c} W_{c c}^p W_{f f}^p 
	\bigg].
\end{eqnarray}
Here the trace is taken over all space-time variables. Terms linear in the applied fields vanish since the net charge densities and currents are identically zero in the ground state. We also omit the spacial dependence of all gauge fields in the long-wavelength approximation.   

\subsubsection{Diamagnetic contributions}

We first deal with the diamagnetic contributions $ G_{c c} W_{c c}^d$ and $ G_{f f} W_{f f}^d$. The calculation for the $c$-term will be shown below.
\begin{eqnarray}
	\Tr \bigg[ G_{c c} W_{c c}^d \bigg]
	&=& \Tr \big[ G_{c c}  \big] \frac{\mathbf{A}_c^2}{ 2 m_c}  
	= \Tr \big[  (\nabla \cdot \mathbf{v}_c)G_{c c} \big] \frac{\mathbf{A}_c^2}{4}  \nonumber \\
	&=& - \frac{1}{4}\Tr \big[
	\mathbf{v}_c \cdot (\nabla G_{c c})  \big]
	\mathbf{A}_c^2  \nonumber \\
	&=& - \Tr 
	 \bigg[ \frac{G_{c c}^2 \mathbf{v}_c^2}{4}  +   V G_{c c} G_{c f} v_c^+ + \frac{G_{cf} G_{fc}\mathbf{v}_c \cdot \mathbf{v}_f }{4}\bigg] \mathbf{A}_c^2 .
\end{eqnarray}
Similarly, for the $f$ part
\begin{eqnarray}
	\Tr \bigg[ G_{f f} W_{f f}^d \bigg] =- \Tr \bigg[
	\frac{G_{f f}^2 \mathbf{v}_f^2}{4}  +  V G_{f f} G_{f c} v_f^- +\frac{G_{cf} G_{fc}\mathbf{v}_c \cdot \mathbf{v}_f}{4}  \bigg] \mathbf{A}_f^2.
\end{eqnarray}

\subsubsection{Paramagnetic contributions}

The paramagnetic contributions are
\begin{eqnarray}
	\Tr \bigg[ \underbrace{\frac{1}{2} (G_{c c} W^p_{c c})^2}_{\mathbf{1}} &+& 
	\underbrace{G_{c c} G_{c f} W^p_{c c} W^p_{f c}}_{\mathbf{2}}  \nonumber 
	+\underbrace{\frac{1}{2} (G_{f f} W^p_{f f})^2}_{\mathbf{1}'}+ \underbrace{G_{f f} G_{f c} W_{f f}^p W^p_{c f}}_{\mathbf{2}'}  \nonumber \\ 
	+\underbrace{G_{c c} G_{f c} W_{c c}^p W_{c f}^p}_{\mathbf{3}} &+& \underbrace{G_{f f} G_{c f} W_{f f}^p W_{f c}^p}_{\mathbf{3}'} + \underbrace{G_{c c} G_{f f} W_{c f}^p W_{f c}^p}_{\mathbf{4}} + \underbrace{G_{c f} G_{f c} W_{c c}^p W_{f f}^p}_{\mathbf{5}} 	
	\bigg] .
\end{eqnarray}

\ew

We will show our calculation term by term.
\begin{eqnarray}
	\mathbf{1} &=& \Tr \bigg[ \frac{1}{2} G_{c c}^2 (i A_c^\tau - \mathbf{v}_c \cdot \mathbf{A}_c)^2 \bigg] ,\nonumber \\
	&=&  \Tr \bigg[ - \frac{1}{2} G_{c c}^2 \bigg](A_c^\tau )^2 + \Tr \bigg[\frac{1}{4} G_{c c}^2 \mathbf{v}_c^2\bigg] \mathbf{A}_c^2 ,\nonumber
\end{eqnarray}
where we use the fact that terms odd under time-reversal or spatial inversion vanish after the trace.  Similarly,
\begin{eqnarray}
	\mathbf{1}' &=& \Tr \bigg[ - \frac{1}{2} G_{f f}^2 \bigg](A_f^\tau )^2 + \Tr \bigg[\frac{1}{4} G_{f f}^2 \mathbf{v}_f^2\bigg] \mathbf{A}_f^2 .\nonumber
\end{eqnarray}
For terms $\mathbf{2}$ and  $\mathbf{2}'$,  
\begin{eqnarray}
	\mathbf{2} &=& \Tr \bigg[
		-V G_{c c} G_{c f} (i A_c^\tau - \mathbf{v}_c \cdot \mathbf{A}_c) A_c^-	
	\bigg]\nonumber \\
	&=& V \Tr \bigg[
		G_{c c} G_{c f} v_c^+	
	\bigg]\mathbf{A}_c^2, \nonumber \\
	\mathbf{2}' &=& V \Tr \bigg[
		G_{f f} G_{f c} v_f^- \bigg]\mathbf{A}_f^2. \nonumber
\end{eqnarray}
For terms $\mathbf{3}$ and  $\mathbf{3}'$,
\begin{eqnarray}
	\mathbf{3} &=& -V\Tr \bigg[ 
		G_{c c} G_{c f} (i A^\tau_c - \mathbf{v}_c \cdot \mathbf{A}_c)A_f^+ \bigg]\nonumber \\
		&=& V\Tr \bigg[ 
			G_{c c} G_{f c} v_c^- 
		\bigg] A_c^- A_f^+,\nonumber \\
	\mathbf{3}' &=& V \Tr \bigg[ G_{f f} G_{c f} v_f^+\bigg] A_c^- A_f^+.\nonumber
\end{eqnarray}
For terms $\mathbf{4}$ and  $\mathbf{5}$,
\bean
		\mathbf{4} &=& V^2 \Tr \bigg[ 
			G_{c c} G_{f f} 	
		\bigg] A_c^- A_f^+ , \nonumber \\
		\mathbf{5} &=& \Tr \bigg[ 
			G_{c f} G_{f c} (i A_c^\tau - \mathbf{v}_c \cdot \mathbf{A})(i A_f^\tau - \mathbf{v}_f \cdot \mathbf{A})	
		\bigg] \nonumber \\
		&=& \Tr \bigg[ 
			G_{cf} G_{fc} \bigg]A^\tau_c A^\tau_f + 
			\Tr \bigg[ 
			G_{cf} G_{fc} \frac{\mathbf{v}_c \cdot \mathbf{v}_f}{2}	\bigg] \mathbf{A_c} \cdot \mathbf{A_f}.
\eean

Collecting all the paramagnetic and diamagnetic terms, we finally get 
\begin{align}
	\frac{\Delta F}{N T} &= 
	-\frac{1}{2} \bigg\{
	\Tr \bigg[ 
		\frac{ \mathbf{v}_c \cdot \mathbf{v}_f}{2} G_{cf} G_{fc}\bigg](\mathbf{A}_c - \mathbf{A}_f)^2   \nonumber\\
		&\hspace{-1em}+ \Tr[G_{c c}^2] (A_c^\tau)^2 - 2 \Tr [G_{cf} G_{fc}] A_c^\tau A_f^\tau+ \Tr[G_{f f}^2] (A_f^\tau)^2 \bigg\} \nonumber \\
		&\hspace{-1em}+ \Tr \left[V^2 G_{c c} G_{f f} + V G_{c c} G_{fc} v_f^+ + V G_{f f} G_{c f} v_c^-\right] A_c^- A_f^+.  
		\label{eq:Higgs1}
\end{align}

\subsubsection{Evaluation of Matsubara sums}

In order to proceed, we conduct the Matsubara summations. For TKI, the Green's function $\mathcal{G}$ can be written as 
\begin{eqnarray}
	\mathcal{G}_\mathbf{k} = \frac{P_\mathbf{k}^l}{z - {E}_{\mathbf{k}}^l} + \frac{P_\mathbf{k}^h}{z - {E}_{\mathbf{k}}^h},
\end{eqnarray}
where $E_{\mathbf{k}}^{l,h}$ are the eigenenergies of the two band system and $P^{l,h}$ are the projection matrix defined as below 
\begin{eqnarray}
	P^{l,h}_{\mathbf{k}} &\equiv&
	\frac{\mathbb{1}}{2} \pm \frac{1}{2 \Delta E_{\mathbf{k}}}\begin{pmatrix}
		\Delta \varepsilon_\mathbf{k} & V k^+ \\
		V k^- & - \Delta \varepsilon_\mathbf{k}
	\end{pmatrix}.
\end{eqnarray} 
Here 
\begin{eqnarray}
	\Delta \varepsilon_\mathbf{k} &=& \frac{k^2}{2 m_c} - 
	{\frac{k^2}{2 m_f}} - (\mu+\lambda),\nonumber \\
	\Delta E_\mathbf{k} &=& E^h - E^l = 
	\sqrt{\Delta \varepsilon_\mathbf{k}^2+(2 kV)^2} .
\end{eqnarray}
Then by doing a Matsubara summation, we have 
\begin{eqnarray}
	T \sum_{n} G_{AB} (i \omega_n) 
	G_{CD}(i \omega_n) =  \frac{P^l_{AB}P^h_{CD} \!+\! P^l_{CD}P^h_{AB}}{E_\mathbf{k}^h - E_\mathbf{k}^l}.\qquad
\end{eqnarray}
Here we assumed that the temperature $T$ is smaller than the energy gap. Hence, we derive the following identities
\begin{eqnarray}
	T \sum_n G_{c c}(i \omega_n) G_{f f}(i \omega_n) &=& \frac{\Delta E^2_\mathbf{k} + \Delta \varepsilon_\mathbf{k}^2}{2 \Delta E^3_\mathbf{k}}, \\
	T \sum_n G_{c c}(i \omega_n) G_{f c}(i \omega_n) &=& - \frac{V k^- \Delta \varepsilon_\mathbf{k}}{2 \Delta E_\mathbf{k}^3},
	\\
	T \sum_n G_{f f}(i \omega_n) G_{c f}(i \omega_n)  &=& \frac{V k^+ \Delta \varepsilon_\mathbf{k}}{2 \Delta E_\mathbf{k}^3}, 
\end{eqnarray}
\begin{align}
	T \sum_n \! G_{c f}(i \omega_n) G_{f c} (i \omega_n)
	&=\!T \sum_n \! G_{f\,f}^2 (i \omega_n)\!=\! T \sum_n \! G_{c c}^2 (i \omega_n)\nonumber\\
	&=\frac{\Delta E_\mathbf{k}^2 - \Delta \varepsilon_\mathbf{k}^2}{2 \Delta E_{\mathbf{k}}^3}.
\end{align}
The last identity implies the coefficients in the second line of Eq.~\eqref{eq:Higgs1} is in a perfect Higgs form.


\subsubsection{Pauli-Villars Regularization} 

Putting the diamagnetic and paramagnetic contributions together, we get 
\begin{eqnarray}
\!\!\frac{\Delta F}{N T} = -\frac{1}{2}\Tr &\bigg[& \,G_{c f} G_{f c}  (A_c^\tau - A_f^\tau)^2 +  \nonumber\\
			& & \frac{ \mathbf{v}_c \cdot \mathbf{v}_f}{2} G_{cf} G_{fc} (\mathbf{A}_c - \mathbf{A}_f)^2  		
\bigg] + F_{1},\qquad
\end{eqnarray}
where the term
\begin{equation}
	F_1 = \left(V^2 G_{c c} G_{f f} + V G_{c c} G_{c f} v_f^+ + V G_{f f} G_{f c} v_c^-\right) A_c^- A_f^+
\end{equation}
superficially violates gauge invariance. To recover gauge invariance, we perform a Pauli-Villars regularization, which is equivalent to adding fictitious bosonic counterparts to the fermionic fields in the free energy to be integrated out:
\begin{eqnarray}
	S_{\mathrm{eff}} &\rightarrow& \int d^d x\, \left(
		\bar{\psi}_f [-\mathcal{G}_f^{-1}] \psi_f + \bar{\phi}_b [-\mathcal{G}_b^{-1}] \phi_b 
	\right),\nonumber \\
	 F_{\mathrm{eff}} &\rightarrow& \Tr \bigg\{\ln [-\mathcal{G}_f^{-1}] - \ln [-\mathcal{G}_b^{-1}]\bigg\}. 
\end{eqnarray}
The bosonic Green's function is defined by changing all the chemical potentials $\lambda$'s to a fictitious chemical potential $\Lambda$. 
\begin{eqnarray}
	\mathcal{G}_b \equiv \mathcal{G}_f(\lambda \rightarrow \Lambda ),\quad \lim_{\Lambda\rightarrow \infty}\mathcal{G}_b = 0. 
\end{eqnarray}
Such a $\Lambda$-cutoff is naturally set on a lattice by the bandwidth of conduction electrons.

We further notice that the difference of chemical potentials of $c$ and $f$ electrons can be absorbed into the temporal component of the gauge field by a gauge transformation.
\begin{equation}
	A_{c,f}^t \rightarrow A_{c,f}^t + \frac{\mu - \lambda}{2}.
\end{equation}
This greatly simplifies our Pauli-Villars regularization calculation. Consequently, we use only a single chemical potential $\lambda$ in all fermionic Green's functions.

After imposing the Pauli-Villars regularization to the term $F_1$ and conducting the Matsubara sum, we are left with a two-dimensional $k$-integral 
\bea
	I &=& \frac{V^2}{2}\int \frac{d^2 k}{(2 \pi)^2}
	\bigg\{ \frac{\left[\Delta\varepsilon^2(\lambda) 
    - 2 \lambda
    \Delta \varepsilon(\lambda) +4(kV)^2 \right]}{\Delta E^3(\lambda)} \nonumber\\
    &&\hspace{1cm}-  \frac{\left[\Delta\varepsilon^2(\Lambda) 
    - 2 \Lambda
    \Delta \varepsilon(\Lambda) +4(kV)^2 \right]}{\Delta E^3(\Lambda)}
	\bigg\},
\eea
where 
\bea
	\Delta \varepsilon(x) &\equiv&
	\frac{k^2}{2 m_c} - \frac{k^2}{2 m_f} - 2x, \nonumber \\ 
	\Delta E(x) &\equiv& \sqrt{\Delta \epsilon^2(x) + 4 k^2 V^2} .
\eea
An elementary but laborious calculation concludes that
\begin{equation}
	I=0.
\end{equation}
Therefore, the effective theory is duly gauge invariant.

\subsubsection{Stiffness and velocity calculation}

In the end, the effective action takes the form
\begin{eqnarray}
	\label{eq:sSeff}
	\frac{\Delta F}{N T} &=&
	- \frac{1}{2}  \bigg[\Gamma^\tau
		( A_c^\tau - A_f^\tau)^2 + 
	   \Gamma^\mathbf{x} (\mathbf{A}_c - \mathbf{A}_f)^2
   \bigg]
	\nonumber \\
	&\equiv&-\frac{1}{2} \Gamma \bigg[
		 ( A_c^\tau - A_f^\tau)^2 + 
		v_\Gamma (\mathbf{A}_c - \mathbf{A}_f)^2
	\bigg],
\end{eqnarray}
where 
\begin{eqnarray}
	\Gamma^\tau =
    2\frac{V^2}{{\pi}} \int_0^\infty {dk}\, \frac{ k^3 }{\Delta E^3}
    =\frac{1}{\pi} \left(\frac{1}{m_c} + \frac{1}{|m_f|}\right)^{-1}\!\!,\;
\end{eqnarray}
and
\begin{eqnarray}
    \Gamma^\mathbf{x} &=&\frac{V^2}{\pi m_c m_f}\int_0^\infty {dk}\, \bigg[
        \frac{ k^5 }{\Delta E^3(\lambda)} - 
        \frac{ k^5 }{\Delta E^3(\Lambda)}  \bigg]\\
        &=& 
        \frac{2 \Lambda}{\pi} \frac{m_c |m_f|}{(m_c + |m_f|)^2} + O(1).
\end{eqnarray}
Thus to leading order,
\begin{eqnarray}
	v_\Gamma =  \frac{2 \Lambda}{m_c + |m_f|}.
\end{eqnarray}
Recall that we used a radial $V$ gauge starting at Eq.\,\eqref{eq:Vreal}. By reversing the gauge fixing on $V$, the effective action \eqref{eq:sSeff} becomes Eq.\,\eqref{eq:Seff}.

\section{Kondo flux repulsion in a single plaquette}\label{sec:paquette}
The basic principle of Kondo flux repulsion can be analyzed on a single Kondo plaquette (Fig.\,\ref{fig1}a). Within large-$N$ MF theory this leads to a 4$\times$4 Hamiltonian. For simplicity, we limit ourselves to particle-hole symmetric case, where $\mu=\lambda=0$ and $\sbraket{c\dg c}=\sbraket{f\dg f}=1/2$ at every site. Since only the flux through the plaquette matters, the phase can always be moved to the hopping between the two $c$ electrons:
\be
{\cal H}\to\matc{cc|cc}{0 & -\abs{t_c}e^{i\alpha} & \abs{V_1} & 0 \\ -\abs{t_c}e^{-i\alpha} & 0 & 0 & \abs{V_2} \\ \hline \abs{V_1} & 0 & 0 & \abs{t_f} \\ 0 & \abs{V_2} & \abs {t_f} & 0},
\ee
This Hamiltonian can be written as
\bea
{\cal H}&=&-t_c\hat\alpha\frac{1+\tau^z}{2}+t_f\sigma^x\frac{1-\tau^z}{2}+\tau^x[\bar V\bb 1+\delta V\sigma^z],\qquad
\eea
where $\vec\tau$ and $\vec\sigma$ are two sets of Pauli matrices and $\hat\alpha=\cos\alpha\,\sigma^x-\sin\alpha\,\sigma^y$. The $\vec\tau$ acts in the space of $c$ and $f$ electrons whereas $\vec\sigma$ acts in the space of sites 1 and 2. Let us further assume $\abs{V_1}=\abs{V_2}$, i.e.\ $\delta V=0$. Squaring $\mathcal{H}$, we get
\be
{\cal H}^2=\Delta{\cal H}^2+\frac{t_c^2+t_f^2}{2}, \quad \Delta{\cal H}^2=A\tau^z+ {\cal B}\tau^x,
\ee
where $A=(t_c^2-t_f^2)/2$ and ${\cal B}=-\bar V[t_c\hat\alpha -t_f\sigma^x]$. Note that
\bea
\abs{{\cal B}}^2=\bar V^2(t_c^2+t_f^2-2t_ct_f\cos\alpha).
\eea
Defining $\hat {\cal B}={\cal B}/\abs{{\cal B}}$, the matrix $\Delta{\cal H}^2$ can be diagonalized with $U=\cos\phi+i\hat {\cal B}\tau^y\sin\phi$. The result is
\be
\Delta\tilde{\cal H}^2=U\dg\Delta{\cal H}^2U= \tau^z\sqrt{A^2+\abs{{\cal B}}^2}.
\ee
This together with the form of $\abs{{\cal B}}^2$ are sufficient to show that the minimization of the free energy w.r.t.\ the Kondo flux $\alpha$ pins $\alpha\to \pi$. Considering the form of the free energy
\be
F=\frac{\abs{V_1}^2+\abs{V_2}^2}{J_K}+\frac{\abs{t_f}^2}{J_H}+E\Big(\abs{V_1},\abs{V_2},\abs{t_f},\alpha\Big),
\ee
we conclude that minimization of free energy w.r.t.\ $\alpha$ is decoupled from its minimization w.r.t.\ other parameters. Therefore indeed $\alpha=\pi$, i.e.\ Kondo flux is repelled.

\section{Mean-field theory in the uniform case}
For the sake of completeness, here we provide the MF analysis of the uniform normal and topological Kondo insulators in absence of the magnetic field. The free energy per unit area ($A$) per spin is
\be
\frac{F}{NA}=-{k_BT}\frac{1}{A}\sum_k\sum_{\tilde r=\pm}\log[1+e^{-\beta E_r(k)}]+\frac{V^2}{J}-\lambda q,
\ee
where $q=Q/N$ and in terms of $\bar\eps_k=(\eps_c+\eps_f)/2$ and $\Delta\eps_k=\eps_c-\eps_f$, the energies are
\be
E_{\pm}(k)=\bar\eps_k\pm \sqrt{(\Delta\eps_k/2)^2+V^2u^2(k)},
\ee
The structure factor $u(k)$ incorporates the spatial structure of the hybridization. For an NKI, $u(k)=1$ whereas for a TKI, $u^2(k)=\sin^2 k_x+\sin^2k_y$. Taking MF w.r.t.\ $V^2$ and $\lambda$ leads to
\bea
&&\frac{1}{A}\sum_k\frac{f[E_-(k)]-f[E_+(k)]}{2\sqrt{(\Delta\eps_k/2)^2+V^2u^2(k)}}u^2(k)=\frac{1}{J},\\
&&\frac{1}{A}\sum_k\frac{f[E_-(k)]-f[E_+(k)]}{2\sqrt{(\Delta\eps_k/2)^2+V^2u^2(k)}}\Delta\eps_k=q-1/2.
\eea
In the following we discuss the MF equations for the continuum limit of NKI and TKI. Doing so for the constraint is slightly subtle but doable \cite{Wugalter20}. Since we would like to study the Kondo breakdown, in the following we focus on the first equation. 

\subsection{NKI}
At $T=0$ using $\veps=\Delta\eps_k/2$, the $1/J_K$ equation becomes
\bea
\frac{1}{J_K}&=&\int_{-D}^{D}\frac{\rho(\veps)d\veps}{2\sqrt{\veps^2+V^2}}, \quad \rho(\veps)=\frac{2k}{2\pi}\frac{dk}{d\veps}
\eea
In a continuum description, $\Delta\eps_k\sim k^2$ and $\rho(\veps)$ is constant
\be
\veps=\frac{k^2}{2m}-\bar\mu, \qquad m=\frac{2m_c\abs{m_f}}{m_c+\abs{m_f}}, \qquad\rho=\frac{m}{2\pi}. 
\ee
where we introduced $\bar\mu=({\mu+\lambda})/2$. For future references, we also define $\Delta\mu=\mu-\lambda$ and observe that
\be
\bar\eps_k=\gamma\frac{k^2}{2m}-\bar\mu, \qquad \gamma=\frac{\abs{m_f}-m_c}{\abs{m_f}+m_c}.
\ee
A reasonable choice of the energy cutoff is $D\sim\bar\mu$. The integral over $\veps$ is logarithmically diverging at UV and we have included a cutoff $D$ to regulate it. At $T\gg T_K$ on the other hand $V=0$ and the integral is also logarithmically diverging at IR. However, the MF equation is only valid at the onset of $V>0$, i.e.\ at $T\le T_K$. The IR cut-off is provided by $T_K$ and we have
\be
\frac{1}{J_K}=\rho\int_{T_K}^D\frac{d\veps}{\veps} \so T_K=De^{-1/\rho J}.
\ee
Alternatively, regarding $J_K(D)$ as the running coupling we can either take a derivative of the previous equation w.r.t.\ $D$, or divide the integration into $(0,D-\delta D)$ and $(D-\delta D,D)$ to get
\be
\frac{dJ_K}{d\ell}=\rho J_K^2, \qquad d\ell=-d \log D,
\ee
which agrees with $J_K$ being a marginally relevant coupling.

\subsection{TKI}

In the continuum limit we can set $u^2(k)=k^2$. At $T=0$, the $1/J_K$ equation becomes 
\bea
\frac{1}{J_K}&=&\int_0^{\Lambda}\frac{d^2k}{(2\pi)^2}\frac{k^2}{2\sqrt{(\Delta\eps_k/2)^2+k^2V^2}},
\eea
where $\Lambda$ is again the UV momentum cut-off and $D_1$ and $D_2$ define the relevant bandwidth around the Fermi energy. The UV divergence in this TKI case is more severe. Additionally there is the usual IR divergence when $\Delta\eps_k=0$. To see this, let us use $\veps=\Delta\eps_k/2$ and assume a flat density of states, so that
\bea
\frac{1}{J_K}&=&\rho\int_{-D_1}^{D_2}{d\veps}\frac{2m(\veps+\bar\mu)}{2\sqrt{\veps^2+2mV^2(\veps+\bar\mu)}},
\eea
which is linearly diverging at UV. Going through the same reasoning, at $T\gg T_K$ we have $V=0$. Therefore,
\bea
\frac{1}{J_K}&=&\rho\int_{-D_1}^{D_2}{d\veps}\frac{2m(\veps+\bar\mu)}{2\abs{\veps}}.
\eea
This integral is linearly diverging at UV, but note that the divergence vanishes if the bandwidth is taken to be symmetric, i.e.\ $D_1=D_2$. Therefore, we conclude that this linear divergence arising from the bandwidth mismatch is not involved in the renormalization of the Kondo coupling and take $D_1=D_2=D$. We also need to include an IR cut off. Finally, we find
\be
\frac{1}{2m\bar\mu J_K}=\rho\int_{T_K}^{D}\frac{d\veps}{\abs{\veps}}.
\ee
If $\bar\mu$ is kept constant, this clearly leads to the same RG flow and $T_K$ as the NKI. If however, $\bar\mu\sim D$ is also varied we can define a new coupling constant $\lambda=2m\bar\mu J_K$ which has the same RG flow:
\be
\frac{d\lambda}{d\ell}=\rho\lambda^2 \quad \rightarrow \quad T_K=De^{-1/\rho\lambda}=De^{-\pi/m J_K}.
\ee
This change in scaling dimension of $J_K$ is due to the non-onsite nature of the Kondo coupling in a TKI which in the continuum limit has the form of $J_K S^b\cdot(\nabla^a c)\sigma^b(\nabla^a c)$.

\section{Lifshitz-Kosevich formula for NI and TI }
\label{AF}
In this section, we first re-derive the LK formula \cite{lifshitz1956} for magnetic oscillations in a metal and then generalize it to the case of both normal and topological band insulators. Similar attempts have been made numerically in \cite{zhang2016quantum} and analytically in \cite{Allocca2022}. Such systems can be described by a 2D Hamiltonain
\be
H=\int{dxdy}\mat{c\\ f}\dg{\cal H}\mat{c\\f}.
\ee
For the normal insulator,
\be
{\cal H}(\vec p)=\mat{\frac{p^2}{2\abs{m_c}}-\lambda & V \\ V & -\frac{p^2}{2\abs{m_f}}+\lambda},
\ee
while for the topological (inverted) insulators (in only one sector),
\be
{\cal H}(\vec p)=\mat{\frac{p^2}{2\abs{m_c}}-\lambda & Vp^- \\ Vp^+ & -\frac{p^2}{2\abs{m_f}}+\lambda}.
\ee
The flat-band case can be reached by taking $m\to\infty$ limit. An overall chemical potential $\mu$ is included later on. 

It is more convenient to use the symmetric gauge in which $(A_x,A_y)= (-y,x)B/2$. Denoting the canonical momenta as $\vec \Pi\equiv \vec p-e\vec A$, a quadratic metal ${\cal H}_0(p)=p^2/2m$ becomes
\begin{equation*}
{\cal H}_0(\vec\Pi)=\frac{1}{2m}(\Pi_x^2+\Pi_y^2)
=\hbar\omega_B(a\dg a+1/2),
\end{equation*}
where $\omega_B=eB/m$ is the cyclotron frequency and we have defined $\Pi^\pm\equiv\Pi_x\pm i\Pi_y$, in terms of which
\be
a\equiv\frac{1}{\sqrt{2m\hbar\omega_B}}\Pi^+, \andd a\dg\equiv\frac{1}{\sqrt{2m\hbar\omega_B}} \Pi^-.
\ee
They obey the usual $[a,a\dg]=1$ algebra. 
The ladder operators have the eigenstates
\be
\hspace{-0.3cm}a\ket{N}=\sqrt{N}\ket{N-1} , \quad a\dg\ket{N}=\sqrt{N+1}\ket{N+1}.
\ee
Therefore, in a normal metal, 
\be
E(N)=(N+\frac{1}{2})\hbar\omega_B,\quad\lr\quad N(E)=-\frac{1}{2}+\frac{E}{\hbar\omega_B}.\nonumber
\ee

At small fields, $N$ is large and $\hbar\omega_B/2$ is unimportant. However, this is not the case for an insulator, and therefore we keep this term. Remarkably, the masses $m$ only affects the energies and not the wavefunction. The latter only depends on magnetic length scale $l_B \! = \! \sqrt{h/eB}$, which is independent of system parameters. Therefore, the orthogonality relations within the Landau levels of the same band extends to in between different $c$ and $f$ bands. Furthermore, defining $m\equiv 2\abs{m_cm_f}/(\abs{m_c}+\abs{m_f})$ and $\gamma\equiv(\abs{m_f}-m_c)/(\abs{m_f}+m_c)$, we can write $E_c(N)=\gamma_c\hbar\omega_B(N+1/2)$ and $E_f(N)=-\gamma_f\hbar\omega(N+1/2)$, where $\gamma_c=m/m_c=1+\gamma$ and $\gamma_f=m/\abs{m_f}=1-\gamma$.

In the case of an NI, we have
\be
H=\sum_{N,k_x}\mat{c_N\\ f_{N}}\dg{\cal H}\mat{c_N\\ f_{N}},
\ee
where the Hamiltonian ${\cal H}$ for an NI becomes
\be
{\cal H}(\vec\Pi)=\mat{\gamma_c\hbar\omega_B(N+\frac{1}{2})-\lambda & V \\ V & -\gamma_f\hbar\omega_B(N+\frac{1}{2})+\lambda}.
\ee
The eigenvalues are
\be
E_\pm(N)=\gamma\hbar\omega_B(N+1/2) \pm\sqrt{[\hbar\omega_B(N+1/2)-\lambda]^2+V^2}.
\ee
For a TI, choosing a slightly different basis we can write
\be
H=\sum_{N,k_x}\mat{c_N\\ f_{N-1}}\dg{\cal H}\mat{c_N\\ f_{N-1}},\label{eqF10}
\ee
where
\be
\hspace{-.35cm}{\cal H}(\vec \Pi)=\mat{\gamma_c\hbar\omega_B(N+\frac{1}{2})-\lambda & \sqrt{2m\hbar\omega_B N}V \\ \sqrt{2m\hbar\omega_B N}V & -\gamma_f\hbar\omega_B(N-\frac{1}{2})+\lambda}.\label{eqF11}
\ee
The eigenvalues are
\bea
\label{eqFF12}
 E_\pm(N)&=&\gamma\hbar\omega_B(N+1/2)\\
 &&\hspace{1cm}\pm\sqrt{[\hbar\omega_B(N+1/2)-\lambda]^2+2m\hbar\omega_B NV^2}. \nonumber
\eea
The form of Eqs.\,\pref{eqF10}, \pref{eqF11} is similar to those in \cite{Grubinskas2018}.

In the grand canonical ensemble, the free energy is
\be
F=-k_BT{\cal N}\sum_{r=\pm 1}\sum_{N=0}^\infty \varphi_r(N). \\
\ee
Here, $r=\pm 1$ and $N$ are the band and LL indices respectively and ${\cal N}={BA e}/{h}$ is the LL degeneracy. The function $\varphi_r(N)$ has the usual form
\be
\varphi_r(N)= \varphi[E_r(N)]=\log[1+e^{-\beta(E_r-\mu)}],
\ee
where an overall chemical potential is included.  Next we use the Poisson summation formula
\bea
\sum_{N=0}^\infty\varphi_r(N)&=&\frac{1}{2}\varphi_r(0)+\int_0^\infty dn\varphi_r(n)\label{eqF16}\\
&&\quad+2{\rm Re}\sum_{\ell=1}^\infty\int_0^\infty{dn}\varphi_r(n)e^{2\pi i\ell n}. \nonumber
\eea
The first two terms are unimportant for quantum oscillations \cite{lifshitz1956}. Concentrating on the third term we can write
\bea
F_3&=&-2k_BT{\cal N}\,{\rm Re}\sum_r\sum_{\ell=1}^\infty \int{dn_r}\varphi_r(n)e^{2\pi i\ell n}.\\ \nonumber
\eea
Integrating by parts here allows us to further simplify the expression since the boundary terms do not contribute to oscillations. Thus,
\bea 
	F_3 &=& 2k_BT{\cal N}\,{\rm Re}\sum_r\sum_{\ell=1}^\infty \int{dn}\frac{1}{2\pi i\ell}\frac{d \varphi}{d n}e^{2\pi i\ell n} \label{eqF17} \\
	&=& 2k_BT{\cal N}\,{\rm Re}\sum_r\sum_{\ell=1}^\infty \int{dE} \frac{d \varphi}{d E} \frac{1}{2\pi i\ell}e^{2\pi i\ell n(E)}
	,\nonumber 
\eea 
where $d \varphi (E)/ d E=-\beta f(E-\mu)$ is related to the occupation of the levels. In the case of a simple metal with parabolic dispersion, $n(E)$ is linear in $E$ and the integral can be done exactly. More generally,  due to the large number of filled Landau levels $n(E)$ near the Fermi energy and its large slope with energy $n'(E)$, the fast oscillatory factor kills the integral at all energies except near the Fermi energy where $f(E-\mu)$ varies. Taylor expanding  $n(E)=n(\mu)+(E-\mu)n'(\mu)$ gives (this is exact for parabolic dispersion),
\be
F_3\approx {\cal N}\sum_{\ell=1}^{\infty}\frac{(-1)^{\ell}k_BT}{\ell\sinh[2\pi^2 \ell n'(\mu)k_BT]}\cos[2\pi\ell n(\mu)],
\ee
where we have analytical continued  the Laplace transform of the Fermi-Dirac distribution ($\tau/\beta<1$ for convergence)
\be
\intinf{\frac{dx e^{x\tau/\beta}}{e^x+1}}=\intoinf{\frac{dy y^{	\tau/\beta-1}}{y+1}}=\frac{\pi}{\sin(\pi\tau/\beta)}.
\ee
Using $n(E)= E/\hbar\omega_B - {1}/{2}$ for a metal, we recover the LK formula
\be
\hspace{-.25cm} F_3\approx{\cal N}\sum_{\ell=1}^{\infty}\frac{(-1)^{\ell}k_BT}{\ell\sinh(2\pi^2 \ell  mk_BT/\hbar eB)}\cos\Big(2\pi\ell\frac{\pi p_F^2}{heB}\Big),
\ee
which has $1/B$ oscillations with a frequency given by the Fermi surface area $\pi p_F^2$, as well as higher harmonics with smaller magnitudes. We can evaluate the two terms dropped in \pref{eqF16}. The first term is
\be
F_1=-\frac{1}{2}k_BT{\cal N}\log[1+\zeta], \qquad {\zeta}\equiv e^{-\beta(\hbar\omega_B/2-\mu)}.
\ee
The second term is
\bea
F_2&=&-k_BT\intoinf{dn}\log[1+\zeta e^{-n\beta\hbar\omega_B}]\nonumber\\
&=&\frac{(k_BT)^2}{\hbar\omega_B}{\rm Li}_2(-\zeta)\approx -\frac{(k_BT)^2}{\hbar\omega_B}\log^2(\zeta),
\eea
where in the last line we assumed $\zeta\gg 1$, i.e.\ $N\gg 1$. Therefore, we confirm that $F_3$ is indeed the only oscillatory part of the Free energy leading to magnetic oscillations.

Generalizing this to an insulator faces the problem that for $E\sim\mu$, the number of Landau levels $n(E)$ is not only small but $n(\mu)=0$. Generally Eqs.\,(\ref{eqF11},\ref{eqFF12}) can be expressed as
\be
E(n)= \gamma \hbar \omega_B\left(n + \frac12 \right) \pm \sqrt{\left[\hbar \omega_B \left(n + \frac12 \right) - \tilde{\lambda}\right]^2 + \Delta^2},
\ee
where 
\begin{align}
\Delta_{\rm NI}&\!=\!V, & \Delta_{\rm TI}&\!=\!\sqrt{mV^2(2\tilde\lambda-\hbar\omega_B+mV^2)},\nonumber\\
\tilde{\lambda}_{\rm NI} &\!=\! \lambda, & \tilde{\lambda}_{\rm TI} &\!=\! \lambda - m V^2.
\end{align}
Essentially, $\Delta$ is related to the band gap. A direct evaluation of Eq.\,\pref{eqF17} leads to
\be
F_3= 2 \mathcal{N} \Re \sum_r \sum_{\ell} J_r(\ell),
\ee
where 
\be 
	J_r(\ell) = -\frac{1}{2 \pi i \ell}  \int_0^\infty dn f[E_r(n)-\mu] \frac{d E_r}{d n}e^{2 \pi i \ell n}.
\ee
For $k_B T \ll \text{gap}$, $f(E_+ - \mu) \to 0$, the contributions mainly come from the occupied bands, i.e., the $r = -1$ branch. $J_-(\ell)$ can be split to two parts $J_r^{(1)}, J_r^{(2)}$, which are defined as 
\bea 
	J_-^{(1)} &=&  \frac{{\gamma \hbar \omega_B}}{2 \pi i \ell} \int_0^{\infty} d n e^{2 \pi i \ell n} \nonumber\\
	&=&\frac{{\gamma\hbar  \omega_B}}{2 \pi i \ell}\lim _{\varepsilon \rightarrow 0^+} \int_0^{\infty} d n e^{(i 2 \pi \ell-\varepsilon) n} = \frac{\gamma \hbar \omega_B}{(2 \pi \ell)^2}, \\
	J_-^{(2)} &=& \frac{{- \hbar \omega_B}}{2 \pi i \ell} \bigintsss_0^{\infty} d n e^{2 \pi i \ell n} \frac{\hbar \omega_B\left(n+\frac{1}{2}\right)-\tilde{\lambda}}{\sqrt{\left(\hbar \omega_B\left(n+\frac{1}{2}\right)-\tilde{\lambda}\right)^2+\Delta^2}}\nonumber \\
	&=&\frac{(-1)^{\ell+1}\Delta}{2 \pi i \ell } e^{2 \pi i \ell\frac{\lambda}{\omega_B}} \int_{x_0}^{\infty} d x \frac{x}{\sqrt{x^2+1}} e^{2 \pi i \ell \frac{\Delta}{\omega_B} x}\\
	&=& \frac{(-1)^{\ell+1}\Delta}{2 \pi \ell i}I_\ell\left({\Delta}/{\omega_B};x_0\right)  \exp(2 \pi i \ell\frac{\lambda}{\hbar \omega_B}) .
\eea 
Here, $I_\ell(\Delta/\omega_B;x_0)$ is defined as Eq.\,\eqref{eq:LKInt} in the main text and $x_0 = (\tilde{\lambda} - \hbar \omega_B/2)/\Delta$.

For a generic $x_0$, the free energy $F$ can only be evaluated numerically.
In the limit $x_0 \to \infty$, a close form for the free energy can be found using 
\bea 
	I_\ell(\Delta/\hbar \omega_B;\infty) = 2 i K_1(2 \pi \ell \Delta/\hbar \omega_B),
\eea 
where $K_1(x)$ is the modified Bessel function of second kind.
This gives 
\bea 
		F = 2  \mathcal{N} \Delta\sum_{\ell=1}^\infty \frac{(-1)^\ell}{\pi \ell} 
	K_1 \! \left(2 \pi \ell \frac{\Delta}{\hbar \omega_B} \right) \cos \! \left( 2 \pi \ell \frac{\tilde{\lambda}}{\hbar \omega_B}\right)\!\!. \quad
\eea 
which is Eq.\,\eqref{eq:F_sol}. This means that the onset of quantum oscillation is at $\hbar \omega_B\sim \Delta$ at low temperature. 

It is also worth mentioning that in the limit $\Delta/\hbar \omega_B \to 0$, due to the asymptotic form of $K_1(x \to 0) \sim 1/x$, one recovers the traditional metallic LK form at zero temperature, 
\bea 
	F =  \mathcal{N}\hbar \omega_B  \sum_{\ell} \frac{(-1)^\ell}{(\pi \ell)^2} \cos \left( 2 \pi \ell \frac{\pi p_F^2}{\hbar \omega_B}\right).
\eea

\bibliography{KL}
\end{document}